\documentclass[12pt,preprint]{aastex}
\usepackage{lscape,morefloats,multirow,color,graphicx}

\begin{document}

\title{Dense Molecular Cores Being Externally Heated}

\author{GWANJEONG KIM\altaffilmark{1,2,6}, CHANG WON LEE\altaffilmark{1,2,7}, MAHESWAR GOPINATHAN\altaffilmark{3}, WOONG-SEOB JEONG\altaffilmark{2,4}, AND MI-RYANG KIM\altaffilmark{1,5}} 

\altaffiltext{1}{Radio Astronomy division, Korea Astronomy and Space Science Institute, 776 Daedeokdae-ro, Yuseong-gu, Daejeon, 34055, Republic of Korea}

\altaffiltext{2}{Department of Astronomy and Space Science, University of Science \& Technology, 217 Gajungro, Yuseong-gu, Daejeon, 34113, Republic of Korea }

\altaffiltext{3}{Aryabhatta Research Institute of Observational Sciences, Manora Peak, Nainital 263129, India}

\altaffiltext{4}{Space Science division, Korea Astronomy and Space Science Institute, 776 Daedeokdae-ro, Yuseong-gu, Daejeon, 34055, Republic of Korea}

\altaffiltext{5}{Department of Astronomy and Space Science, Chungbuk National University, 52 Naesudong-ro, Heungdeok-gu, Cheongju Chungbuk, 28644, Republic of Korea }

\altaffiltext{6}{e-mail:archer81@kasi.re.kr}
\altaffiltext{7}{corresponding author}

\begin{abstract}
We present results of our study on eight dense cores, previously classified as starless, using infrared (3-160 {\micron}) imaging observations with \textit{AKARI} telescope and molecular line (HCN and N$_2$H$^+$) mapping observations with \textit{KVN} telescope. Combining our results with the archival IR to mm continuum data, we examined the starless nature of these eight cores. Two of the eight cores are found to harbor faint protostars having luminosity of $\sim0.3-4.4$ L$_{\odot}$. The other six cores are found to remain as starless and probably are in a dynamically transitional state. The temperature maps produced using multi-wavelength images show an enhancement of about 3-6 K towards the outer boundary of these cores, suggesting that they are most likely being heated externally by nearby stars and/or interstellar radiation fields. Large virial parameters and an over-dominance of red asymmetric line profiles over the cores may indicate that the cores are set into either an expansion or an oscillatory motion, probably due to the external heating. Most of the starless cores show coreshine effect due to the scattering of light by the micron-size dust grains. This may imply that the age of the cores is of the order of $\sim10^{5}$ years, being consistent with the timescale required for the cores to evolve into an oscillatory stage due to the external perturbation. Our observational results support the idea that the external feedback from nearby stars and/or interstellar radiation fields may play an important role in the dynamical evolution of the cores.
\end{abstract}

\keywords{ISM: clouds {\textemdash} ISM: kinematics and dynamics {\textemdash} ISM: individual (CB22, CB246-2, L1041-2, L1234, L1512, L1517B, L1582A, and L1621-1)}

\section{Introduction}

Starless cores are dense molecular regions of interstellar medium with density of a few 10$^{4}$ cm$^{-3}$ and temperature of $\sim$10 K where there are no embedded Young Stellar Objects (YSOs) \citep[e.g.,][]{Myers:1983jg, Myers:1983cw, WardThompson:1994tg, Lee:1999fi, Bergin:2007iy}. They are believed to be the potential sites of future star formation \citep[e.g.,][]{Lee:1999fi, Bergin:2007iy}.

For a better understanding of star formation process, it is essential to have a better knowledge of how these cores form and evolve. According to the current understanding, the cores are thought to form basically in filamentary clouds and evolve as a result of fragmentation through gravitational instability when their masses exceed the critical mass per unit length. However, in reality the details of how various processes such as ambipolar diffusion and turbulence dissipation work in their formation are still under debate \citep{Shu:1987dp, Ciolek:1995kt, Nakano:1998db, Myers:1998ee, Andre:2014et}. Further, the final fate of the dense cores depends on the initial conditions set by various physical environment in which they reside. Thus, the formation process of the dense cores may not be as simple as we understand. However, it is certain that out of the entire starless cores, the sufficiently dense cores have to undergo gravitational collapse to initiate star formation in them \citep[e.g.][]{Lee:2011dx}.

In order to understand the current evolutionary status of the starless cores, it is important to study their dynamical properties. One way of performing this is to carry out spectroscopic observations using molecular lines which show asymmetry with self-absorption in the line profiles of optically thick species (e.g. CS 2-1, HCN 1-0) and a Gaussian profile in optically thin species (e.g. C$^{18}$O 1-0, N$_2$H$^+$ 1-0). In previous studies, based on single pointing observations toward the central region of starless cores, it was seen that approximately one fourth of the starless cores show a double peak profile with blue asymmetry (the blue peak is higher in intensity than the red peak) indicative of inward motion \citep{Walker:1994ip, Lee:1999de, Lee:2004bi, Sohn:2007gn}. However, of the remaining cores, a number of them displayed red asymmetry, which is the reverse of the blue asymmetry, suggestive of expansion motion. In more extensive mapping observations which covered the entire projected area of the starless cores, roughly half of them showed the dominance of blue asymmetric profiles over the whole mapped core area \citep{Lee:2001jm}. This result suggests that these cores are undergoing gravitational infall and about to begin star formation. However, some of the remaining cores which include B68 and FeSt 1-457, showed only red asymmetry or complex patterns of blue and red asymmetry, indicative of expansion or oscillation motions, respectively \citep{Lee:2001jm, Lada:2003ec, Redman:2006cu, Aguti:2007jq}. Moreover, these line profiles sometimes change their asymmetries according to the molecular line tracers in which the observations are made. This is because of the fact that different molecular lines or transitions have different optical depths and trace different region of the dense cores \citep{Gregersen:1997uc}.

Compiling the results from previous molecular line surveys, \citet{Lee:2011dx} made a statistical analysis to investigate the evolution of starless cores in various environments. Based on their study they suggested that the starless cores may evolve through a number of distinguishable evolutionary stages by increasing their column density. To begin with, the cores evolve through a static stage which is then followed by an expanding and/or an oscillating stage. The star formation eventually occurs through the final stage of global contraction motion once the cores become adequately dense ($\rm N\gtrsim 6\times10^{21}~cm^{-2}$).

In this paper we intend to address the issue of how a dense core can evolve from static stage to an oscillating and/or expanding stage. For this, we have selected eight dense cores that are classified as starless and are found to be located in different environments like isolated (CB22, CB246-2, L1512, L1234) or associated with filamentary cloud (L1041-2, L1517B, L1621-1) or HII region (L1582A). From previous molecular line observations, all these cores are known to show blue/red asymmetry \citep{Lee:1999de, Lee:2001jm, Lee:2004bi, Sohn:2007gn}. We made continuum imaging observations of these cores using \textit{AKARI} satellite and radio spectroscopic observations of the entire projected area of the cores in HCN (1-0) and N$_2$H$^+$ (1-0) molecular lines. The main aim of the \textit{AKARI} observations is to re-examine the starless status of the cores by looking for any point source(s) embedded in them. We used data from 1.2-1200 {\micron} to construct the spectral energy distribution (SED) of YSO candidates, if found, to understand their nature. We combined our \textit{AKARI} observations and data sets from 2MASS, \textit{Spitzer}, \textit{WISE}, \textit{JCMT}, and \textit{IRAM} to construct the SEDs. In two of the eight cores we found point sources showing characteristic properties of YSOs. Then we also constructed temperature maps of all the eight cores by combining continuum images from \textit{AKARI} and other longer wavelength continuum images obtained from the archive (Table \ref{tbl:data}). This was done to look for any temperature variation over the cores. Finally, using radio observations in HCN and N$_2$H$^+$ lines we studied the current dynamical status of these cores. We describe the source selection and observations in Section 2. We show the results based on the infrared images and molecular line profiles, and discuss the results and the relation between them in Section 3. The conclusions of this study are summarized in Section 4.

\section{Observations \& Data Reduction}

\subsection{Source Selection \& \textit{AKARI} Observations} 

We made observations of eight dense cores using \textit{AKARI} satellite. These sources were previously classified as starless based on the non-detection of any embedded sources in the Infrared Astronomical Satellite (IRAS) observations \citep{Lee:1999fi}. However, spectral line profiles observed in these cores showed some characteristic features indicative of internal motions related to the initial star-forming processes. The line profiles seen in the eight cores are: a blue asymmetry in L1582A (in HCN 1-0; \citealt{Sohn:2007gn}), a red asymmetry in CB246-2 (in CS 2-1 and CS 3-2; \citealt{Lee:1999de,Lee:2001jm,Lee:2004bi}), CB22, L1234, L1512, L1517B, and L1621-1 (in HCN 1-0; \citealt{Sohn:2007gn}), and a mixture of blue and red asymmetry in CB246-2 and L1041-2 (in HCN 1-0; \citealt{Sohn:2007gn}). FoV of the \textit{AKARI} maps was well suited for the typical sizes of our sample starless cores. Relatively high sensitivity and wide wavelength coverage of the \textit{AKARI} enabled us to observe the main part of the SED of any faint point sources that are present within the FoV. Table \ref{tbl:gen} lists some of the basic parameters of our eight target cores along with the source identification numbers (OBSID) and dates on which \textit{AKARI} observations were carried out.

The \textit{AKARI} was an infrared space telescope of 68.5 cm diameter launched by the Japan Aerospace Exploration Agency (JAXA) on 2006 February 21 \citep{Murakami:2007uw,Nakagawa:2007vk}. The telescope was equipped with two focal-plane instruments: an Infrared Camera (IRC; \citealt{Kaneda:2007uw,Onaka:2007tt,Tanabe:2008um}) and Far-Infrared Surveyor (FIS; \citealt{Kawada:2007tm,Shirahata:2009ti}). The imaging devices of the IRC had three channels of NIR (1.8-5.3 {\micron}), MIR-S (5.4-13.1 {\micron}), and MIR-L (12.4-26.5 {\micron}). The FIS imaging instrument had two wide bands (WIDE-S and WIDE-L) having central wavelengths at 90 and 140 {\micron} and two narrow bands (N60 and N160) with central wavelengths of 65 and 160 {\micron}. The FoVs for the IRC and the FIS were 10{\arcmin}$\times$10{\arcmin} and 17{\arcmin}$\times$14{\arcmin}, respectively. The pointing error of the telescope was better than 3{\arcsec} \citep{Murakami:2007uw}. 

The \textit{AKARI} observations for our sources were carried out at 3, 4, 7, and 11 {\micron} in the IRC02 mode and at 65, 90, 140, and 160 {\micron} in the FIS01 mode between 2006 September and 2007 July. In the IRC02 mode the images were obtained with two fixed filters in one pointed observation for an area of 10{\arcmin}$\times$10{\arcmin} while in the FIS01 mode the images were taken in two round-trips with a cross-scan shift of 8 arcsec sec$^{-1}$ for an area of 17{\arcmin}$\times$14{\arcmin}. The IRC and FIS data were retrieved using the OBSID of each object from the DARTS archive\footnote{DARTS is maintained by C-SODA at ISAS/JAXA. See http://darts.isas.jaxa.jp/astro/akari/}. The data reduction was carried out using the IRC imaging pipeline version 20091022 and the FIS Slow-Scan Toolkit Version 20070914, respectively. Observation parameters and the sensitivities of our \textit{AKARI} observations are summarized in Table \ref{tbl:akari}.

\subsection{Molecular Line Observations}

Although we selected our target cores based on the spectral line features, these features were identified based on the molecular line observations made towards the central regions of the cores only. Also, some of the observations lacked the signal-to-noise (S/N) ratio required to properly characterize the observed line shapes. Therefore we carried out mapping observations for all our target sources in HCN (1-0) and N$_2$H$^+$ (1-0) molecular lines with three 21-m radio telescopes of Korean VLBI Network (KVN) which are located in Seoul (Yonsei site), Ulsan (Ulsan site), and Jeju island (Tamna site) of South Korea. The HCN (1-0) molecular line is known to be an optically thick tracer for gas motions in the density of $\rm \sim 10^5~cm^{-3}$ and shows various asymmetry shapes in the three hyperfine lines with different relative opacities under the local thermal equilibrium condition \citep[e.g.,][]{Sohn:2007gn}. The N$_2$H$^+$ (1-0) molecular line is considered as an optically thin tracer for the motion of gas in the densest region ($\rm \gtrsim 10^6~cm^{-3}$) of the core and consists of seven hyperfine lines \citep[e.g.,][]{Tafalla:2006dw}.

 These observations were carried out in a single dish observing mode between 2012 December and 2014 May. The data were obtained in a frequency switch mode with a frequency offset of 4 MHz with dual (left and right circular) polarizations. The two polarization profiles were averaged to get a single spectrum with a better S/N ratio. As a back-end instrument, we used an autocorrelation spectrometer with a 32 MHz bandwidth and 7.825 kHz spectral resolution (corresponding to a velocity resolution of $\sim 0.026$ km s$^{-1}$ in HCN line and $\sim 0.025$ km s$^{-1}$ in N$_2$H$^+$ line). The beam sizes (FWHM) and the main beam efficiencies of the three telescopes at Yonsei, Ulsan, and Tamna were 32{\arcsec}-33{\arcsec} and 0.31-0.33 at 93 GHz, respectively (from http://kvn.kasi.re.kr). The on-source integrated time for each position was typically between 15 and 30 minutes to achieve a S/N ratios of $\gtrsim$10 in both HCN and N$_{2}$H$^{+}$ lines at the central region of the cores, with the typical system temperature of 200-300 K at both frequencies during the observations. The pointing of the telescope was checked on an hourly basis using SiO maser sources located in the close vicinity of our target cores and found to be better than 4{\arcsec}. The data reduction like folding, baseline-subtracting, and averaging of the molecular line spectrum was carried out using the CLASS software of the GILDAS package\footnote{http://www.iram.fr/IRAMFR/GILDAS/}. Table \ref{tbl:kvn} presents the observed size and the sensitivities of the mapped regions of our target sources in both the molecular lines. 

\subsection{2MASS, \textit{Spitzer}, \textit{WISE}, \textit{JCMT}, \& \textit{IRAM} Data}

Apart from our \textit{AKARI} and molecular line observations, we also compiled multi-wavelength (IR to mm) images and photometric information for the eight cores from various data archives. The data from the Two Micron All Sky Survey (2MASS), the Spitzer Space Telescope (Spitzer), and the Wide-Field Infrared Survey Explorer (WISE) were obtained from the NASA/IPAC Infrared Science Archive\footnote{http://irsa.ipac.caltech.edu/}. Because the 2MASS and the \textit{WISE} missions were all sky surveys, we obtained data at 1.2, 1.7, 2.2, 3.4, 4.6, 12, and 22 {\micron} for all the eight cores. However, \textit{Spitzer} observations of the full extend of the cores at 3.6, 4.5, 5.8, 8.0, and 24 {\micron} are available only CB22, L1512, and L1517B. L1582A and L1041-2 were observed only partially by \textit{Spitzer} at the same wavelengths mentioned above. CB246-2 was observed at 3.6, 4.5, 5.8, and 8.0 {\micron}, but not at 24 {\micron}. The \textit{Spitzer} and the \textit{WISE} missions had an overlap in their wavelength coverages. In situations where the data from both these facilities were available, we preferred to use the one from the \textit{Spitzer} due to its better angular resolution and sensitivity compared to the \textit{WISE}. For 22 {\micron} data, we used the data from the \textit{WISE} mission.

The continuum data at millimeter or sub-millimeter regime are very important to constrain the physical properties of dense cores and embedded sources better. However, the data in these wavelength regime are available only for five out of eight sources in either data archives or previous publications. We obtained 850 {\micron} data for three of our cores, namely CB246-2, L1512 (fully observed), and L1517B (partially observed) from the JCMT science archive\footnote{http://www.cadc.hia.nrc.gc.ca/jcmt/search/product/}. We also used continuum 1200 {\micron} data for L1041-2 and L1582A from \citet{Kauffmann:2008jj}. Table \ref{tbl:data} summarizes the availability of continuum data for our target cores. 

\subsection{Photometry}

Even though all the cores selected by us were classified previously as starless, some of them were found to contain faint point sources. In order to study them in detail and discern their true nature, we performed aperture photometry on these detected point sources. The aperture photometry was made using the IDL\footnote{http://www.exelisvis.com/ProductsServices/IDL.aspx} routine `aper.pro', particularly for the point sources detected at wavelengths longer than 20 {\micron} as we were interested in only those sources that are embedded and currently forming inside the core. The point sources were extracted in cases where the signal was about 3$\sigma$ above the background noise at 22, 24, or 65 {\micron}. The positions of the point sources were obtained from the 3.6 {\micron} images. Using the IDL routine, gcntrd.pro, we computed the coordinates of the point sources by performing Gaussian fits to their emission distribution. The photometry was carried out on all the sources detected in images of all the wavelengths except for those from 2MASS and \textit{WISE} bands. The photometric results in 2MASS and \textit{WISE} bands are already available in the point source catalogs \citep{Cutri:2003vra,Cutri:2012tx}. The results from our photometry were compared with those from 2MASS and \textit{WISE} database. The results were found to be in good agreement within the uncertainty. The aperture radius and the sky annulus (inner radius and width) used for the aperture photometry were set to be different at different wavelengths in order to avoid any contamination from the neighbouring sources.

In the IRC02 mode of \textit{AKARI}, the observations in the 3 and 4 {\micron} were obtained with two exposures of 44.4 seconds and 4.7 seconds, respectively. The long exposure observations were made to detect the faint sources while the short exposure observations were made to avoid the saturation of the bright sources in the frame. The \textit{AKARI} images with the long-exposures sometimes showed artifacts like muxstripe, muxbleed, and column pulldown basically due to the presence of a bright point source. For sources affected by these artifacts, we used the short-exposure image for the photometry. The measured flux density from the point sources was converted into the physical unit `Jansky' by applying the conversion factor and the aperture correction factor for each band \citep{Engelbracht:2007ec,Hora:2008he,Tanabe:2008um,Shirahata:2009ti}. 

\section{Results \& Discussions}

\subsection{Multi-wavelength Images of Eight Dense Cores}

The \textit{AKARI} images of the eight cores studied here are presented in Figure 1. Images of the cores in other wavelengths, obtained from the archives, are also presented in the same Figure which is ordered according to the increasing wavelength. Starting with the optical images in the top left panel to the 1200 {\micron} in the bottom right panel, the central wavelength corresponding to each of the images is labelled on the top left corner. A contour on each of the panels is a half minimum contour level of the extinction region in the optical image, indicating the approximate boundary of a dense core. Presence of dense cores is conspicuous in all the optical images as dark opaque patches. The cores appear dark because the submicron-sized solid particles present in them extinct the light coming from the background stars. However, because of the less extinction suffered in near infrared wavelengths, the light from the background stars can penetrate through the opaque regions of the dense cores. This makes the cores transparent in the wavelength range of $\sim1.2-\sim4.5$ {\micron}. As a result numerous background point sources are visible through the cores. Some of these sources could be embedded protostars. It is interesting to note that six of the eight dense cores in our sample show some scattered diffuse emission at 3.0-4.5 {\micron} wavelength range. This is thought to be due to ``core-shine effect'' \citep{Pagani:2010ik}. More discussion on this is presented in section 3.3.2. 

The number of point sources seen toward the dense cores is drastically reduced between 5.8 and 11 {\micron} and the scattered light due to the core-shine effect also appears to diminish. Instead, at these wavelengths, the background becomes bright with diffuse emission and the dense cores appear dark. This effect gets more prominent at 11 {\micron}. This is believed to be due to the emission coming from the polycyclic aromatic hydrocarbon (PAH) situated along the Galactic plane in the background \citep{Tielens:2008fx}. The PAH emission features peak at 7.6 and 11.3 {\micron} and thus are slightly shifted from the central wavelengths of IRC bands. However, IRC bands at 7 and 11 {\micron} are wide enough (1.8 and 4.1 {\micron}, respectively; \citealt{Onaka:2007tt}) to include some of the strong PAH emission features and hence collect emission originating from them. Most of the cores in our sample are seen in absorption with respect to the bright PAH background emission. 

Point sources visible at $22-65$ {\micron} are believed to be good candidates of embedded YSOs \citep[e.g.,][]{Robitaille:2006cb, Robitaille:2007dl, Whitney:2013cw}. While no point sources with characteristic properties of YSOs are detected within the observed fields of \textit{AKARI} towards six of our target cores, a few YSO candidates are found in two of them, namely L1582A and L1041-2. Figures \ref{fig:multiL1582A} and \ref{fig:multiL1041-2} show that there are eight possible YSO candidates in L1582A and two in L1041-2, respectively. The heating of the surrounding dust by the radiation from these embedded YSOs is believed to be the main reason for the higher brightness at 90, 140, and 160 {\micron}. Other dense cores (CB22, L1517B, L1512, L1621-1, L1234, and CB246-2) also show some emission at these wavelengths particularly near the boundaries of the cores, even though there is no evidence of any heating source within them. We believe that such emission is caused by the heating of the cores externally. Properties of the detected YSO candidates and their effects on the parent cores are discussed in detail in the next section. The effects of external heating on the starless cores in our sample are discussed further in the section 3.3.

\subsection{Newly Detected Sources Embedded in ``Starless'' Dense Cores}

We found a number of point sources, likely to be embedded, in two of the dense cores which were previously classified as starless by \citet{Lee:1999fi}. The detected point sources are believed to be protostars because of their higher brightness at 22-65 {\micron} than at the shorter wavelengths \citep[e.g.,][]{Robitaille:2006cb, Robitaille:2007dl, Whitney:2013cw}. In this section we derive their bolometric temperature ($T_{bol}$) and bolometric luminosity ($L_{bol}$) and explore whether they are associated with the cores or not.

\subsubsection{Bolometric Temperature and Luminosity of the Point Sources}

 The $T_{bol}$ and $L_{bol}$ are useful quantities to investigate the physical properties of YSOs \citep[e.g.,][]{Myers:1993en, Chen:1995eo}. These quantities can be estimated from the SED of the sources \citep[e.g.,][]{Myers:1993en, Evans:2009bk}. The $T_{bol}$ is defined as the temperature of a blackbody having the same mean frequency $\nu$ as the observed continuum spectrum \citep{Myers:1993en} and can be derived by the equation:
\begin{equation}
T_{bol}=[\zeta(4)/4\zeta(5)]h\bar{\nu}/k=1.25\times10^{-11}\bar{\nu}~\rm KHz^{-1},
\end{equation}
where $\zeta(n)$ is the Riemann zeta function of argument $n$, $h$ is the Planck's constant, and $k$ is the Boltzmann's constant \citep{Myers:1993en}. Here the mean frequency $\bar{\nu}$ is obtained from the ratio of the first and the zeroth frequency moments of the spectrum \citep{Ladd:1991ez}:
\begin{equation}
\bar{\nu}=\frac{\int^{\infty}_{0}\nu S_{\nu}d\nu}{\int^{\infty}_{0} S_{\nu}d\nu}.
\end{equation}
On the basis of the calculated $T_{bol}$ values, point sources can be classified into five evolutionary groups, which are \citep[e.g.,][]{Evans:2009bk}:
\[\begin{array}{l}
T_{bol} < 70~K: Class~0 \\
70~K\leq T_{bol} \lesssim 350~K: Class~I \\
350~K\lesssim T_{bol} \lesssim 950~K: Flat~spectrum\\
950~K\lesssim T_{bol} \leq 2800~K: Class~II \\
2800~K< T_{bol} : Class~III \\
 \end{array}
\]
The $L_{bol}$, the total energy emitted by the point source per unit time, is calculated by integrating the flux over the full observed SED, assuming that the source has a spherical geometry \citep{Chen:1995eo};
\begin{equation}
L_{bol}=4 \pi D^{2} \int^{\infty}_{0}S_{\nu} d \nu,
\end{equation}
where $D$ is the distance from the observer to the core and $S_{\nu}$ is the flux density at the specific frequency. We derived $T_{bol}$ and $L_{bol}$ for all the point sources detected in L1582A and L1041-2. A good coverage of their SEDs from 1.2 {\micron} to 1.2 mm (as shown in Figure \ref{fig:sed}) enabled us to estimate $T_{bol}$ and $L_{bol}$ quite accurately. Based on the values of $T_{bol}$, we found that all of them are likely to be protostars of Class 0, I, or the flat spectrum sources. Their luminosities are found to be in the range 0.3$\sim$4.4 L$_{\sun}$. The derived bolometric temperature, bolometric luminosity, classification, and the flux densities of these protostellar candidates are summarized in Table \ref{tbl:flux}. We note that the values of $L_{bol}$ that are obtained in this work could be significantly affected by the uncertainty due to the unknown geometry of the Class I and the flat spectrum sources. Because of the difficulty in determining the geometry of these sources due to the lack of knowledge of parameters such as the inclination of the disk and the cavity, the opening angle of the cavity and so on, we ignored this uncertainty. Therefore the uncertainty of $L_{bol}$ quoted by us was obtained by simply propagating the errors of the distance and the flux densities in equation 3. The uncertainty of the values of $T_{bol}$ was also obtained by propagating the errors of the flux densities. The errors due to the aperture photometry are the main source of uncertainty in the measurements of the flux densities.

\subsubsection{Association of Point Sources with Dense Cores}

It is important to ascertain whether the YSO candidates identified by us in L1582A and L1041-2 are actually associated with them or not. It could be possible that they are simply a chance projection along the line of sight and not physically related to the cores at all. However, if they are related, then their presence could modify some of the properties of the cores like the temperature structure and the kinematics of the material surrounding them. 

Using the photometric data at 90, 140, 160, 850, and 1200 {\micron}, we created the temperature maps of all the cores studied here. The available data for the each of the dense cores are listed in Table \ref{tbl:data}. We first subtracted the sky background intensity by selecting an emission free region close to our target cores, re-gridded to the same pixel scales (15{\arcsec}), and convolved to 41{\arcsec} beam which corresponds to the lowest angular resolution among the 90-1200 {\micron} images to look for any variation in the temperature in the vicinity of the identified YSO. The temperature corresponding to each set of pixels was estimated by fitting the flux densities obtained in those pixels at 90-1200 {\micron} with the modified blackbody function given by 
\begin{equation}
S_{\nu} (\nu)= \Omega (1-e^{-\tau(\nu)})(B_{\nu}(\nu,T_d)-I_{bg}(\nu))
\end{equation}
with
\begin{equation}
\tau(\nu) = N_H m_H \frac{M_d}{M_H} \kappa_d (\nu),
\end{equation}
where $S_{\nu}(\nu)$ is the flux density at frequency $\nu$, $\Omega$ is the solid angle, $\tau(\nu)$ is the optical depth, $B_{\nu} (\nu,T_d)$ is the Planck function, $T_d$ is the dust temperature, $I_{bg}$ is the background intensity, $N_H=2\times N(H_2)+N(H)$ is the total hydrogen column density, $m_H$ is the hydrogen mass, $M_d/M_H$ is the dust-to-hydrogen mass ratio, and $\kappa_d(\nu)$ is the dust mass absorption coefficient \citep[e.g.,][]{Nielbock:2012ih,Launhardt:2013cr}. In equation 5, $\kappa_d(\nu)$ is obtained from the tabulated values given by \citet{Ossenkopf:1994tq} for mildly coagulated composite dust grains with thin ice mantles called ``OH5.'' Although $I_{bg}$ is dominated by the cosmic background radiation and the diffuse Galactic background, we did not consider them separately since our sky subtraction procedure would remove their contributions also. Equation 4 for various values of temperature is fitted to the observed fluxes of each set of pixels. The temperature corresponding to the fit with the least ${\chi}^2$ value is considered as the temperature of the corresponding pixels. Five of the eight dense cores have data at wavelengths (either 850 or 1200 {\micron}) beyond the expected peak of their SED. Because of this reason we could produce temperature maps quite reliably for L1517B, L1512, L1582A, L1041-2, and CB246-2. However, the remaining three dense cores lack continuum data at these longer wavelengths. Therefore, our estimated value of the temperature is considered as an upper limit. Despite this, the temperature maps produced without using longer wavelengths are still useful to examine the presence of any spatial temperature variation in them. Our temperature maps for the cores with embedded sources are shown in Figure \ref{fig:rgb1}.

We also made two sets of color-composite (CC) images for all the eight cores. The first one is made using 7, 22, and 65 {\micron} and the second using 65, 90, and 160 {\micron}. The CC image of a core produced using higher angular resolution images in wavelengths shortward of 65 {\micron} is useful to detect any embedded point sources. On the other hand, the CC map generated using the images in wavelengths longward of 65 {\micron} is useful to detect the cold (T$\lesssim$25 K) envelopes of embedded sources. As the Planck function ratio 160 to 65 {\micron} at the dust temperature of $\lesssim$25 K is significantly larger than unity, our CC images (with blue for 65 {\micron}, green for 90 {\micron}, and red for 160 {\micron}) are expected to show red color in the immediate vicinity of the embedded sources unless the dust temperature of the envelope is as high as about 50 K from where the Planck function ratio 160 to 65 {\micron} is lower than unity.

In L1582A, the point sources identified as YSO candidates are clearly seen in the CC image made using 7, 22, and 65 {\micron} images. All but IRS 3a/b are located all along the southern and western edges of the core. The brightest of them, L1582A-IRS6, is identified with a known pre-main sequence star V453 Ori \citep{Dolan:2001ey}. Our $L_{bol}$ estimation is quite consistent with the value calculated by \citet{Cohen:1979cc}. There are a number of Herbig-Haro sources detected in the vicinity of V453 Ori and L1582A-IRS5 suggesting that the region is an active site of current star formation \citep{Magakian:2004ima}. The CC image at 65, 90, and 160 {\micron} of L1582A shows enhanced emission at the longer wavelength of 160 {\micron} at the locations of the YSO candidates identified in L1582A, being its color as red. This implies that the YSO sources in L1582A are surrounded by dust envelopes which are not heated as high as about 50 K but remain cold. The temperature map for L1582A does not show any clear local temperature enhancement at the location of most YSOs except for L1582A-IRS4 and IRS7. This may indicate that the YSOs are not luminous enough to heat the entire dense core or the expected local heating of their immediate environment is most likely smoothed out by the large beam (41{\arcsec}) of observations. One noticeable feature seen in L1582A is that the temperature is found to be enhanced along the southern and western edges of the core. The warm south-western boundary may actually indicate local heating by L1582A-IRS4 and 7, which seem less embedded than the other YSOs such as L1582A-IRS2 and 3. These two YSOs may thus heat the dense core from this side. There may be another possibility for the local heating in the south-western edge of L1582A. L1582A is a part of the bright-rimmed cloud, Barnard 30, and its southern edge is pointing roughly towards the $\lambda$ Ori cluster (Collinder 69) which includes the O8 III$+$B0 V binary $\lambda$ Ori AB \citep{Duerr:1982jn,Zhou:1988jj,Zhang:1989wc,Lang:2000vr}. It is possible that L1582A, located $\sim20$ pc away from the central star $\lambda$ Ori, is heated externally causing the temperature towards the southern and western edges to rise. At the moment it is not clear which of the two effects dominates. From CC images and temperature maps, it is not certain that the detected most YSO candidates are physically related to L1582A.
 
In L1041-2, we detected two possible YSO candidates, IRS1 and IRS2, which could be embedded in the core. The CC image with 65, 90, and 160 {\micron} shows red color around two YSO candidates, implying that they are also surrounded with cold envelopes like the case of L1582A. However, at the location of L1041-2-IRS2, which is located just outside the boundaries of the dense core, the CC map also shows a local bluish spot, and the corresponding temperature map displays more clearly a local temperature enhancement, again indicating that this YSO is less embedded and the heating of its immediate surroundings can therefore detected. However, there is no obvious local enhancement in the temperature around L1041-2-IRS1. This may be similar to the cases of most YSOs in L1582A. Moreover, enhancement in the temperature is seen towards the eastern and northeastern parts of the core which is thought to be due to the external heating by the interstellar radiation field. In L1041-2, using CC maps and temperature maps, it also seems to be difficult to determine the physical association between YSOs and their parent clouds.

However, it is still generally acceptable to suggest that the YSOs are physically associated with their parent cores L1582A and L1041-2, simply from the spatial correlation between the YSOs and the high extinction of the dense core which would probably prohibit seeing a background source through it, and the statistical improbability of a by-chance-alignment of the YSOs and the cores (L1582A and L1041-2).

Investigation of the kinematic structure of the dense cores is another way to ascertain the physical association of the YSO candidates with them. For this reason, we made mapping observations of the dense cores using N$_2$H$^+$ (1-0) to look for any line broadening around the envelopes due to star-forming activities. We obtained line widths of N$_2$H$^+$ (1-0) by making seven hyperfine Gaussian fit to the lines. The (D) panels of Figure \ref{fig:rgb1} show the variation of line width in contour around the candidate protostars in L1582A and L1041-2. We noticed that the line widths are significantly broadened around the locations of the YSO candidates in L1582A (IRS 3a/b) by a factor of $\sim 1.5$ and in L1041-2 (IRS1 and IRS2) by a factor of $1.6\sim 2.2$ when compared with the line widths of the other positions in the cores. We also made mapping observations of the cores using HCN (1-0) line to look for any evidence of infall asymmetry in these cores. Figure \ref{fig:HCNmap1} shows the line profile maps of the main component (F=2-1) among three HCN hyperfine lines superposed with the half maximum contour of the N$_2$H$^+$ integrated intensity. It is noted that most of them show a blue asymmetry in double peaks where blue peak is brighter than the red peak, indicative of inward motions of gas in the dense core. The kinematic signatures obtained above also suggest that the YSO candidates detected in these cores are most likely associated with the cores.

\subsection{``Starless'' Dense Cores}

\subsubsection{External Heating of the Starless Cores and Its Effects}

In the previous section we found that two of the eight dense cores, formerly classified as \textit{starless} based on the non-existence of IRAS point sources \citep{Lee:1999fi}, are now identified as small clouds with multiple embedded (proto-stellar) cores. However, the other six dense cores in our sample are still starless. Here we discuss their physical status and possible evolution in future by using continuum and line emission data. 

Like the two proto-stellar cores discussed in the previous section, we present CC images and temperature maps for six starless cores in Figure 5. In the CC images at 7, 22, and 65 {\micron}, the absence of any bright point source at 22 and/or 65 {\micron} indicates that these dense cores are indeed starless. We found that the brightness of our starless cores tends to be more centrally concentrated at longer wavelengths. L1517B, L1512, and CB246-2 show high brightness towards their central region at either 850 or 1200 {\micron} as shown in Figure 1, implying that they are centrally more condensed with cold gas and dust material. However, we could not examine the cloud structure of the other three starless cores, CB22, L1621-1, and L1234, due to the lack of data towards the longer wavelengths. In contrast, all the six cores were observed and detected with strong emission at 90, 140, and 160 {\micron}. Especially the cores CB246-2 and L1517B show bright emission towards the central region at both 140 and 160 {\micron}. This implies that their dust emission is concentrated more towards the central region as manifested in both 140 and 160 {\micron} and in 850 or 1200 {\micron} wavelength regimes. But, the emission pattern seen in the other four starless cores, namely CB22, L1512, L1621-1, and L1234 are somewhat different. In these four cores, the emission from 90 to 160 {\micron} is brighter towards the outer regions of the cores where the dust column density is relatively lower compared with that towards the central regions. Notably, the temperature maps produced for these cores (Figure 5) show an increase of 3$\sim$6 K towards the outer parts of the cores when compared with the central regions. One interesting feature seen in the wavelength range from 90-160 {\micron} in six of the eight cores is the asymmetric emission (and also temperature) pattern enhanced towards a particular portion of the outer rim of the cores. This effect has been noticed in other dense cores also, for example, B68 \citep{Nielbock:2012ih}, CB130 (with 11 dense cores, \citet{Launhardt:2013cr}), L1155, and L1148 \citep{Nutter:2009fk}. 

Heating by the isotropic interstellar radiation field (ISRF) and/or nearby stars was considered as the possible reason for the observed asymmetric rim heating. \citet{Nutter:2009fk} carried out the radiative transfer modelling of a dense core having a Plummer-like sphere with both the isotropic ISRF and directional nearby stellar radiation field, showing that the increase (3-6 K) in the temperature toward the boundaries of the core can be reproduced by the mean ISRF and the asymmetric emission feature could be produced by the presence of any nearby exciting star(s) at the IR wavelength bands. We believe that while the mean ISRF alone may cause an observable increase in the temperature towards the outer parts of the cores, presence of nearby single or multiple stars, in addition to the ISRF heating, as a second order effect, may cause heating of the side of the cores which is facing the star(s).

In order to examine this further, we made a search for any potential source or sources around the six cores that could externally heat and hence enhance the temperature towards the boundaries. The search was conducted by visually inspecting the nearby stars whose spectral type and distance are known from the Simbad astronomical database and the Hipparcos catalogue \citep{Nesterov:1995vx,vanLeeuwen:2007dc}. In at least three of the six cores, namely CB22, L1517B, and L1512 we found potential candidates which are situated within an angular separation of $1^{\circ}$ and towards the side where we noticed the enhancement in the temperature in the cores. These heating sources are found to be of spectral types A or F and are 0.2-0.7 pc away in projected distance from the dense cores. For the remaining three cores (L1234, L1621-1, and CB246-2) we could not find any potential star, whose distance is known, in the vicinity. However, in the case of L1234 we find that a very bright IRAS point source is located 2.2{\arcmin} away from it and to the side where the maximum heating in the core is noticed. From the SED of the IRAS source, we found that this object is a YSO of Class II and bright enough (L$\rm_{bol} \sim$ 47 L$_{\sun}$) to heat the southern part of L1234, provided that it is located at a distance similar to that of the core. 

In the case of L1621-1, an increase in the temperature is noticeable to the northern part of the core. Interestingly, similar increase in the temperature (towards the northern part) is also visible in another core located to the south of L1621-1. This provides convincing evidence for the existence of a possible exciting source to the northern direction of L1621-1. We searched for potential candidates to the northern direction of L1621-1 and found several sources such as HD 39952 (A3 type) and a number of IRAS sources. In CB246-2, the temperature map shows an enhancement all around the boundary of the core, implying that multiple sources situated around it may be responsible for the observed temperature profile. We found a number of potential candidates of A to F type stars within 10 arc minutes angular distance of CB246-2. More information on the potential exciting candidates found towards L1621-1 and CB246-2 are required to confidently determine the source or sources responsible for the external heating of these cores. In Table \ref{tbl:origin}, we list all the potential sources that could possibly heat the respective cores externally with their properties like the spectral type, distance, and their projected distance from the core. In Figure \ref{fig:dss_sw_w} ($1^{\circ}\times1^{\circ}$ DSS optical images) we show their spatial locations with respect to the cores. The cores are delineated using half minimum contour of optical extinction regions in yellow. The curve in red indicates the direction of the core where the envelopes show a relative enhancement in temperature.

Another way to diagnose whether a core has been subjected to any external affect is to look for an asymmetric shape in a molecular line profile toward the cores. \citet{Lada:2003ec}, \citet{Redman:2006cu}, and \citet{Aguti:2007jq} have shown that any disturbance to the core by an external pressure possibly due to, for instance, nearby OB stars can create oscillatory motions in the cores. This could produce spatially complex pattern of red and blue asymmetry line profiles over the dense core depending on the oscillation mode of the core. In Figure 7 we show profile maps of F=2-1 components of HCN (1-0) lines overlaid on the half maximum contour of the integrated intensity of N$_2$H$^+$ (1-0) lines for our six starless cores. Four of the starless cores, namely L1512, L1517B, L1621-1, and CB246-2 display the dominance of ``red'' asymmetry line profiles, i.e., the double peaks where blue peak is fainter than the red peak. CB22 is found to consist of two $\rm N_2H^+$ sub-cores. While its northern sub-core shows a dominance of blue asymmetric profiles over the entire half maximum contour of $\rm N_2H^+$ integrated intensity, the southern core exhibits a mixture of blue and red profiles. This result shows that the two sub-cores are in different kinematic status. The northern sub-core is in overall inward motion whereas the southern sub-core is experiencing a complex oscillatory motion. In the case of L1234, its kinematic behaviour is unclear because, although the HCN line appears asymmetric, its peak velocity coincides with that of $\rm N_2H^+$. To summarize, the kinematics inferred from HCN (1-0) line profiles for all the starless cores, except for L1234, is surprisingly consistent with what is expected from the perturbation by the external heating of the dense cores. Such oscillation motions can cause the cores to increase their central density to change their dynamical status to contracting motions \citep{Stahler:2009ev}. Therefore we believe that these cores are under external perturbation and may be in dynamically unstable status either in a combination of expansion and contraction mode or in an expansion-dominant mode. 
 
Alternative way to diagnose whether the cores are under the effect by the external heating and their stability is in a consistency with the overabundance of the ``red'' asymmetric line profiles toward the cores is to conduct a Virial analysis. 
In order to do this analysis, we need to derive virial and core masses. By ignoring the effects of external pressure, magnetic field, and rotation, the virial mass of a core is given by 
\begin{equation}
M_{vir} = \frac{5}{8 ln2}\frac{R \Delta {v_m}^2 }{\alpha \beta G},
\end{equation}
where $R$ is the radius of a dense core, $\Delta v_m$ is the Full Width at Half Maximum (FWHM) of the line profile of a gas with a mean molecular mass, $G$ is the gravitational constant, and $\alpha$ is the geometric factor for eccentricity. The value of $\beta$ is the correction factor for a non-uniform density distribution ($\beta=(1-p/3)/(1-2p/5)$), where $p$ is the power-law index of the density profile ($\rho \propto r^{-p},~0\leq p \leq 2$) \citep[e.g.,][]{McKee:1992ei,Caselli:2002cb,Chen:2007iy,Miettinen:2012hm}. 
The $\Delta v_m$ is obtained from the molecular line observation here in $\rm N_2H^+$ by 
\begin{equation}
\Delta {v_m}^2 = \Delta {v_T}^2 + \Delta {v_{NT}}^2 = \Delta {v_{obs}}^2 + 8ln2\frac{kT_k}{m_H}(\frac{1}{\mu}-\frac{1}{\mu_{obs}}),
\end{equation}
where $\Delta v_T$ is the thermal line width, $\Delta v_{NT}$ is the non-thermal line width, $\Delta v_{obs}$ is the FWHM of the observed molecular line, $k$ is the Boltzmann constant, $T_k$ is the kinetic temperature, $m_H$ is the mass of the hydrogen atom, $\mu$ is the mean particle weight (=2.33 amu), and $\mu_{obs}$ is the mass of the observed molecule (=29 amu) \citep{Fuller:1992bn}. In the calculation of the virial mass of our six starless cores, we simply assume that the cores are uniform spherical spheres ($\alpha=1$, $p=0$) with a kinetic temperature of 10 K \citep[e.g.,][]{Benson:1989bn,Caselli:2002cb}. The observed line width $\Delta v_{obs}$ is obtained with an average of the FWHM line widths for the $\rm N_2H^+$ line profiles within the half maximum contour of $\rm N_2H^+$ intensity distribution.

The core mass can be calculated by using a simple equation $M = \mu_{H_2} m_H N_{H_2} \Omega_{A}$, where $\mu_{H_2}$ is the mean molecular weight per hydrogen molecule ($\mu_{H_2}$=2.8; \citet{Kauffmann:2008jj}), $m_H$ is the mass of a hydrogen atom, $N_{H_2}$ is the column density of hydrogen molecules, $\Omega_{A}$ is the area of a dense core. The area ($\Omega_{A}$) enclosed within the half maximum of the integrated intensity of N$_2$H$^+$ (1-0) molecular line is considered in the calculation. Because the dense cores are not ideally spherical, we define an effective radius ($r=\sqrt{A/\pi}$) of the area as the dense core radius ($R$). The column density ($N_{N_2H^+}$) can be calculated by using the equation (A4) of \citet{Caselli:2002eg}. To get the column density, we need to measure the integrated intensity ($I$) and the excitation temperature ($T_{ex}$) from molecular line data. The integrated intensity is given by integrating over a velocity range of seven hyperfine line components. The excitation temperature ($T_{ex}$) can be obtained by fitting hyperfine structures of N$_2$H$^+$ (1-0) line into its seven components. So the mass of a dense core can finally be derived from the column density of H$_2$ by adopting the abundance of X$(\rm N_2H^+)\sim$ 6.8 $(\pm4.8)\times10^{-10}$ as derived by \citet{Lee:2011dx} for 35 starless cores from the table provided by \citet{Johnstone:2010ik}. The mass of the cores estimated based on the N$_2$H$^+$ line intensity is given in the column 7 of Table \ref{tbl:virial}.

In some case the abundance of N$_2$H$^+$ molecule can be chemically affected depending on the environments where the cores exist and thus the mass of the cores using N$_2$H$^+$ line observations can be highly uncertain. Therefore we also derived the core mass ($\rm M_{A_V}$) using 2MASS extinction map as another mass measurement. From the extinction maps produced based on the 2MASS star counts \citep{Dobashi:2011ua}, we measure the total extinction value (in magnitude) within the half maximum contour of the N$_2$H$^+$ intensity map and then convert it to the H$_2$ column density of core by using the relation of $\rm N(H_2)/A_V = 9.4\times10^{20}~cm^{-2}~mag^{-1}$ \citep{Bohlin:1978dw}. The calculated masses of the cores are given in the column 8 of Table \ref{tbl:virial}. We compare the masses based on the N$_2$H$^+$ line and those using the 2MASS extinction. Both the estimates of the core masses are found to be consistent within the calculation uncertainties.

For all the cores in our sample, we estimated the ratio between the virial mass and the core mass which is defined as $\alpha = 2K/|W| = M_{vir}/M_{obs}$, where $K$ is the internal kinetic energy of a dense core, $W$ is the gravitational energy, $M_{vir}$ and $M_{obs}$ are the virial and the core mass, respectively \citep[e.g.,][]{Bertoldi:1992kf,McKee:1992ei,Kauffmann:2013bi}. On the basis of the value of $\alpha$, we can diagnose the dynamical instability of a dense core. The $\alpha >1$ or $\alpha <1$ may mean that the dense core would either expand or collapse, respectively. The $\alpha \approx 1$ may mean that the dense core would be under stable condition and in virial equilibrium. Observed and virial masses for all the six starless cores including their related quantities are listed in Table \ref{tbl:virial}. Interestingly enough, although the masses have large uncertainties, all the cores have a virial parameter ($\alpha$) larger than unity. There are many uncertain factors in the mass estimation for the virial parameters. We discuss these to ascertain the significance level of the virial values estimated by us.

The uncertainties in the factors such as the distances of the cores and the measurements of the molecular line widths may cause the virial mass estimation quite uncertain. However, among these, the uncertainty caused due to our assumption of a constant density distribution for the cores is thought to be the dominant one. For example, if the cores have the power-law index p of $\sim 2$ instead of having a constant density distribution (p=0) as assumed here, the virial mass would decrease by $\sim40\%$. The uncertainties of the virial masses given in Table \ref{tbl:virial} are those estimated by propagating the errors of the parameters in equation 6.

The uncertainty in N$_2$H$^+$ abundance contributes more to the uncertainty of the core mass estimated using the N$_2$H$^+$ data than the errors introduced due to the uncertainty in the measurement of the integrated intensity and the distance of the cores. One of the major factors that makes the estimation of the abundance uncertain is the chemical differentiation of the N$_2$H$^+$ \citep[e.g.,][]{Busquet:2010by}. The value of abundance adopted by us could make the core mass estimation uncertain by a factor of $\sim 1.4$.

The core mass estimated using the 2MASS extinction maps, $\rm M_{A_V}$, can also be uncertain due to a number of factors such as the noise error in the extinction values and the uncertainty in the distance to the core. In addition to this, the $\rm M_{A_V}$ values can also be systematically underestimated since the high column density regions of the dense cores can be selectively missed due to the possible lack of stars towards these regions. We tested this possibility by estimating the masses based on the $\rm A_V$ values from the 2MASS extinction maps of the dense cores for which the masses have already been estimated using 1.2 mm continuum data \citep{Kauffmann:2008jj}. The core masses estimated based on the $\rm A_V$ values from the 2MASS extinction maps are found to be underestimated by about 40$\%$ which is considered to be the dominant error in the estimation of $\rm M_{A_V}$ values.

In this manner we derive all possible uncertainties in the mass estimation and in the final virial parameters for the cores by propagating the errors. The virial parameters for all the starless cores are found to be significantly larger than the unity (considering their uncertainties). This implies that all the six cores are, most likely, either expanding or are confined by the external pressure.

In summary, the far-IR emission and temperature enhancement in the boundary of the dense cores, the overdominance of asymmetric ``red'' profiles in HCN line over the cores, and virial parameters larger than unity of the cores are all consistent with the suggestion that our dense starless cores are externally heated and being under unstable status, i.e., expanding or confined by the external pressure.

\subsubsection{Coreshine effects in Dense Cores and Its Implication}

A coreshine effect, first reported by \citet{Pagani:2010ik} with an analogy of the cloudshine discovered in large clouds by \citet{Mathis:1977hp} and \citet{Foster:2006bv}, occurs when the central parts of dense cores appear bright due to the scattered lights from micron-size dust grains that are buried deep inside the cores. This is shown with a contrasting feature between the scattered emission at 3 and 4 {\micron} and the absorption at 7 and 11 {\micron}, and better identified when there exists no embedded source inside the cores to radiate at the near infrared wavelengths as illustrated in Figure \ref{fig:cutprofile1}. Three of the six starless cores (CB22, L1517B, and L1512) in our sample are found to show the coreshine effect in the most prominent contrast between the scattered emission and absorption at the above two wavelength regimes. The coreshine effect is also present in two other starless cores, L1621-1 and CB246-2 but appears weak partly because of the contamination due to the scattered light from the background. On the other hand, the starless core, L1234, does not clearly show any coreshine effect. One possible explanation for the absence of coreshine effect in L1234 may be that the age of this core is relatively lower than the other cores and hence may not have gotten enough time to produce the micron-size dust grains. However the detection of N$_2$H$^+$ and a fairly high column density ($\rm \sim 4\times10^{21}~cm^{-2}$) toward the central region of L1234 indicate that this possibility is highly unlikely as the $\rm N_2H^+$ is known to be a good tracer of quiescent dense gas in the later stages of prestellar cores \citep[e.g.,][]{Bergin:2007iy}. Another explanation for the absence of coreshine effect in L1234 is the shattering of dust grain due to nearby supernovae scenario. It has been suggested that the cores residing in the vicinity of a supernova may not show such coreshine effect due to the possible destruction of the micron-size dust grains by the blast waves coming from the supernova \citep{Pagani:2012dc}. Upon a search carried out by us, we found a supernova remnant, SNR G112.0-01.2, in the close (0.7$^{\circ}$) vicinity of L1234 \citep{Kovalenko:1994wx}, making us to speculate that such a supernova activity might be responsible for the absence of coreshine effect in it.

In any case this effect is important in the sense that it can give an evidence for the existence of large (micrometer-size) dust grains which are believed to be formed through the continuous grain growth process which takes place over the evolution of interstellar clouds. For this reason, the presence of coreshine could be used as a good indicator of the age of the cores \citep{Pagani:2010ik, Steinacker:2010iv}. Numerical calculations of the grain coagulation from the typical grain-size distribution in the interstellar medium by \citet{Hirashita:2013fw} have shown that it would take at least several free-fall time (a few $10^{5}$ years) to form micrometer-size dust grains in a number density of 10$^5$ cm$^{-3}$. Therefore, the dense cores showing the coreshine effect may not be dynamically transient, but likely to be long-lived objects. The life time inferred from the coreshine effect is consistent with the life time inferred from the dynamical status of our starless cores. \citet{Lee:2011dx} have shown that starless cores in different internal motions may reflect their different stages of evolution in terms of the increasing column density. In the previous section we found that our starless cores are likely to be in expanding and/or oscillating mode according to their HCN spectral shapes and large virial parameters. Then their age, if they are in a dynamical status, is suggested to be $\sim 3 \times 10^{5}$ years \citep{Lee:2011dx} which is interestingly comparable to the coagulation time of micron-size grain that produces the coreshine effect in the cores.

We note that four starless cores (CB22, L1517B, L1512, and CB246-2) in our sample have the higher column densities (listed in the column 6 of Table \ref{tbl:virial}) than the critical ones ($\rm \sim 6\times10^{21}~cm^{-2}$) over which the cores are likely to be contracting \citep{Lee:2011dx}. We believe that these four cores are most likely to be in the verge of collapse to initiate star formation in them.

\section{Conclusions}

We present results of our imaging observations of eight, previously classified, starless dense cores using \textit{AKARI} telescope. Using \textit{KVN} telescope, we also carried out mapping observations of these cores in HCN and N$_2$H$^+$ lines. Combining our results with those from the data obtained from other space/ground-based facilities in multi-wavelength (near-IR - millimeter), we re-examined the starless nature of these cores. Of the eight cores studied here, two of them, L1582A and L1041-2, were found to harbor point sources showing properties of young stellar objects. The luminosities of these YSO candidates are in the range from 0.3-4.4 L$_{\odot}$. Absence of any embedded source in the remaining six cores confirmed their starless nature. The temperature maps produced for these six starless cores using multi-wavelength continuum images show enhancement of about 3-6 K towards the outer boundaries which suggests that these cores are getting heated externally. We found a few potential A-F type stars and/or interstellar radiation fields in the vicinity of some of these cores which could be responsible for heating these cores externally. The optically thick HCN (1-0) line profiles in five of the six starless cores (CB22S, L1517B, L1512, L1621-1, and CB246-2) show a dominant red asymmetry over their projected area. This indicates that the cores are perturbed due to the external radiation and are set into an expansion or oscillatory motion. The results from the virial analysis also are found to be consistent with the expansion or oscillatory status.

Five of the starless cores studied here show evidence of coreshine effect along with the temperature enhancement towards the outer boundary and red profiles in their molecular maps. Considering that the coreshine effect is produced due to the scattering of radiation by the micron-size dust grains present deep inside the cores, the age of the cores (of the order of 10$^{5}$ years) inferred from the coagulation time of the grains is found to be consistent with the timescale required for the cores to attain oscillatory stage of evolution. This suggests that these cores are not transient but are long-lived dense cores. Four out of these five cores (CB22, L1517B, L1512, and CB246-2) are found to have the column densities higher than the critical ones ($\rm \sim 6\times10^{21}~cm^{-2}$) and may be close to the collapse stage from which stars can form. Our observational results altogether suggest that the external feedback by nearby stars and/or interstellar radiation fields may play a significant role in the dynamical evolution of the cores.

\acknowledgments
We thank the anonymous referee for giving us important suggestions and comments that have improved the quality of the manuscript significantly. This work was supported by Basic Science Research Program though the National Research Foundation of Korea (NRF) funded by the Ministry of Education, Science, and Technology (NRF- 2013R1A1A2A10005125). This research is based on observations with AKARI, a JAXA project with the participation of ESA. We are grateful to all staff members in KVN which is a facility operated by KASI. This research has made use of the NASA/IPAC Infrared Science Archive, which is operated by the Jet Propulsion Laboratory, California Institute of Technology, under contract with the National Aeronautics and Space Administration. The SIMBAD database, operated at CDS, Strasbourg, France, were also used for this study. 



\clearpage

\begin{deluxetable}{ccccccccc}
\tabletypesize{\scriptsize}
\tablewidth{0pt}
\rotate
\tablecaption{Availability of Multi-wavelength (1.2-1200 {\micron}) continuum data from the observations by \textit{AKARI} and other facilities}
\tablehead{Source name & 2MASS & IRC & IRAC & WISE & MIPS & FIS & SCUBA & MAMBO \\
& (1.2, 1.7, 2.2 \micron) & (3, 4, 7, 11 \micron) & (3.6, 4.5, 5.8, 8.0 \micron) & (22 \micron) & (24 \micron) & (65, 90, 140, 160 \micron) & (850 \micron) & (1200 \micron) 
}
\startdata
CB22 & O & O & O & O & O & O & X & X \\
L1517B & O & O & O & O & O & O & O\tablenotemark{a} & X \\
L1512 & O & O & O & O & O & O & O & X \\
L1582A & O & O & O & O & O\tablenotemark{a} & O & X & O \\
L1621-1 & O & O & X & O\tablenotemark{a} & X & O & X & X \\
L1041-2 & O & O & O\tablenotemark{a} & O & O & O & X & O \\
L1234 & O & O & X & O & X & O & X & X \\
CB246-2 & O & O & O & O & X & O & O & X
\enddata
\tablenotetext{a}{The data are available for the partial region only of the core.}
\label{tbl:data}
\end{deluxetable}

\begin{deluxetable}{cccccccccc}
\tabletypesize{\scriptsize}
\tablewidth{0pt}
\tablecaption{AKARI observations for the eight dense cores}
\tablehead{Object & R.A.\tablenotemark{a} & Dec.\tablenotemark{a} & Distance & Ref.\tablenotemark{b} & \multicolumn{4}{c}{AKARI observations\tablenotemark{c}} \\
\cline{6-9}
name & (J2000) & (J2000) & (pc) & & IRC & obs. date & FIS & obs. date}
\startdata
CB22 & 04:40:39.9 & +29:52:59 & 140$\pm$20 & 1 & 3160007 & 2007 Mar. 04 & 3160008 & 2007 Mar. 03\\
L1517B & 04:55:18.8 & +30:38:04 & 140$\pm$20 & 1 & 3160009 & 2007 Mar. 07 & 3160010 & 2007 Mar. 07 \\
L1512 & 05:04:09.7 & +32:43:09 & 140$\pm$20 & 1 & 3160011 & 2006 Sep. 11 & 3160012 & 2007 Mar. 08 \\
L1582A & 05:32:03.4 & +12:31:05 & 400$\pm$40 & 2 & 3160055 & 2007 Mar. 14 & 3160056 & 2007 Mar. 13 \\
L1621-1 & 05:55:59.4 & +02:18:02 & 500$\pm$140 & 3 & 3160013 & 2007 Mar. 20 & 3160014 & 2007 Mar. 19 \\
L1041-2 & 20:37:17.8 & +57:49:21 & 440$\pm$40 & 4 & 3160041 & 2006 Dec. 14 & 3160042 & 2006 Dec. 14 \\
L1234 & 23:17:57.6 & +62:26:39 & 300$\pm$50 & 5 & 3161083 & 2007 Jul. 23 & 3161084 & 2007 Jul. 24 \\
CB246-2 & 23:56:49.2 & +58:34:29 & 140$\pm$20 & 6 & 3160051 & 2007 Jan. 23 & 3160052 & 2007 Jan. 23 \\
\enddata
\tablenotetext{a}{The coordinates of targets are from the Table 2 of \cite{Lee:1999fi}.}
\tablenotetext{b}{References for the adapted distance: (1) \citealt{Elias:1978ir}; (2) \citealt{Murdin:1977um}; (3) \citealt{Maddalena:1986bu}; (4) \citealt{Felli:1992ws}; (5) \citealt{Lee:1999fi}; (6) \citealt{Snell:1981hg}}
\tablenotetext{c}{Observing identification number (`OBSID') and dates of \textit{AKARI} observation with IRC and FIS, respectively.}
\label{tbl:gen}
\end{deluxetable}

\begin{deluxetable}{ccccc}
\tabletypesize{\scriptsize}
\tablewidth{0pt}
\tablecaption{Specifications of \textit{AKARI} instruments}
\tablehead{Instrument & Wavelength & Bandwidth & FWHM\tablenotemark{a} & Sensitivity (5$\sigma$)\tablenotemark{b}  \\
& (\micron) & (\micron) & ($\arcsec$) & (mJy) }
\startdata
\multirow{4}{*}{IRC} & 3 & 0.9 & 4.0 & 0.016 \\
    & 4 & 1.5 & 4.2 & 0.016 \\
    & 7 & 1.8 & 5.1 & 0.074 \\
    & 11 & 4.1 & 4.8 & 0.132 \\
\hline
\multirow{4}{*}{FIS} & 65 & 21.7 &32.0 & 110\\
    & 90 & 37.9 & 30.0 & 34.0 \\
    & 140 & 52.4 & 41.0 & 350.0 \\
    & 160 & 34.1 & 38.0 & 1350.0 \\
\enddata
\tablenotetext{a}{The FWHMs of IRC and FIS are from \citet{Onaka:2007tt} and \citet{Shirahata:2009ti}, respectively.}
\tablenotetext{b}{The sensitivities of IRC and FIS are from \citet{Onaka:2007tt} and \citet{Kawada:2007tm}, respectively.}
\label{tbl:akari}
\end{deluxetable}

\begin{deluxetable}{ccccc}
\tabletypesize{\scriptsize}
\tablewidth{0pt}
\tablecaption{Molecular line observations}
\tablehead{Object & \multicolumn{2}{c}{Mapping Size} & \multicolumn{2}{c}{Sensitivity} \\
\cline{2-5}
 & HCN & N$_{2}$H$^{+}$ & $\rm \sigma_{{T_A}^{\ast}(HCN)} [K]$ & $\rm \sigma_{{T_A}^{\ast}(N_2H^+)} [K]$}
\startdata
CB22    & 330$\arcsec\times$300$\arcsec$ & 240$\arcsec\times$180$\arcsec$ & 0.09 & 0.10 \\
L1517B  & 180$\arcsec\times$210$\arcsec$ & 150$\arcsec\times$210$\arcsec$ & 0.12 & 0.14\\
L1512   & 240$\arcsec\times$270$\arcsec$ & 180$\arcsec\times$210$\arcsec$ & 0.08 & 0.13 \\
L1582A  & 240$\arcsec\times$180$\arcsec$ & 210$\arcsec\times$120$\arcsec$ & 0.07 & 0.09 \\
L1621-1 & 60$\arcsec\times$120$\arcsec$ & 60$\arcsec\times$120$\arcsec$ & 0.09 & 0.08 \\
L1041-2 & 180$\arcsec\times$240$\arcsec$ & 150$\arcsec\times$240$\arcsec$ & 0.08 & 0.08 \\
L1234   & 120$\arcsec\times$120$\arcsec$ & 90$\arcsec\times$60$\arcsec$ & 0.06 & 0.06 \\
CB246-2 & 240$\arcsec\times$270$\arcsec$ & 180$\arcsec\times$150$\arcsec$ & 0.09 & 0.10 
\enddata
\label{tbl:kvn}
\end{deluxetable}

\begin{deluxetable}{ccccccccccc}
\tabletypesize{\scriptsize}
\rotate
\tablewidth{0pt}
\tablecaption{YSOs detected in two dense cores}
\tablehead{Object & Coordinates & $\rm L_{bol}$\tablenotemark{a} & $\rm T_{bol}$\tablenotemark{b} & Classification &
\multicolumn{6}{c}{Fluxes\tablenotemark{c}} \\
\cline{6-11}
& & & & & 1.24 $\mu m$ & 1.66 $\mu m$ & 2.16 $\mu m$ & 3.0 $\mu m$ & 3.6 $\mu m$ & 4.0 $\mu m$ \\
& & & & & 4.5 $\mu m$ & 5.8 $\mu m$ & 7 $\mu m$ & 8.0 $\mu m$ & 11 $\mu m$ & 22 $\mu m$ \\
& & & & & 24 $\mu m$ & 65 $\mu m$ & 90 $\mu m$ & 140 $\mu m$ & 160 $\mu m$ & 1200 $\mu m$ \\
\cline{6-11}
& (J2000) & (L$_{\sun}$) & (K) & & \multicolumn{6}{c}{(mJy)}}
\startdata
L1582A-IRS1 & 05:32:12 +12:29:42.0 &    1.6$\pm$0.5 & 63$\pm$20 & 0 & 0.31$\pm$0.02 & 0.73$\pm$0.06 & 1.9$\pm$0.2 & 1.5$\pm$0.1 & 2.4$\pm$0.1 & 3.0$\pm$0.1  \\
                            & & & & & 3.6$\pm$0.2 & 4.1$\pm$0.2 & 9.6$\pm$1.1 & 7.4$\pm$0.3 & 14$\pm$1 & 290$\pm$65  \\
              & & & & &  \nodata  & 1132$\pm$46 & 1899$\pm$54 & 13501$\pm$321 & 10300$\pm$200 & 110$\pm$3  \\
\hline 
L1582A-IRS2 & 05:32:08 +12:30:34.6 &    2.8$\pm$0.8 &  97$\pm$26 & I & 0.32$\pm$0.03 & 1.6$\pm$0.1 & 6.0$\pm$0.5 & 18$\pm$1 & 23$\pm$1 & 43$\pm$1  \\
                        & & & & & 40$\pm$1 & 50$\pm$1 & 66$\pm$2 & 48$\pm$1 & 48$\pm$2 & 245$\pm$55  \\
              & & & & &  \nodata  & 2561$\pm$67 & 4361$\pm$80 & 18244$\pm$373 & 18174$\pm$266 & 479$\pm$7  \\
\hline 
L1582A-IRS3a & 05:32:03 +12:31:17.4 &    1.2$\pm$0.2 & 124$\pm$21 & I & 0.32$\pm$0.02 & 3.1$\pm$0.2 & 8.2$\pm$0.6 & 16$\pm$2 & 13$\pm$1 & 15$\pm$1  \\
                         & & & & & 13$\pm$1 & 12$\pm$1 & 15$\pm$1 & 12$\pm$1 & 14$\pm$1 & 160$\pm$3  \\
           & & & & & 80$\pm$3 & 739$\pm$26 & 1410$\pm$33 & 7320$\pm$167 & 11280$\pm$148 & 299$\pm$4  \\
\hline 
L1582A-IRS3b & 05:32:03 +12:31:05.5 &    2.0$\pm$0.6 & 406$\pm$108 & Flat & 0.97$\pm$0.07 & 14$\pm$1 & 57$\pm$4 & 94$\pm$1 & 138$\pm$1 & 127$\pm$1  \\
                   & & & & & 142$\pm$1 & 122$\pm$1 & 118$\pm$2 & 101$\pm$1 & 120$\pm$2 & 311$\pm$70  \\
          & & & & & 272$\pm$5 & 865$\pm$28 & 1676$\pm$35 & 8270$\pm$178 & 11280$\pm$148 & 335$\pm$4  \\
 \hline
L1582A-IRS4 & 05:31:54 +12:31:33.4 &    0.9$\pm$0.3 &  71$\pm$26 & I & 0.52$\pm$0.04 & 1.4$\pm$0.1 & 2.4$\pm$0.2 & 4.2$\pm$0.1 & 1.8$\pm$0.1 & 3.4$\pm$0.1  \\
                               & & & & & 2.4$\pm$0.1 & 2.9$\pm$0.2 & 4.8$\pm$0.6 & 4.5$\pm$0.2 & 4.0$\pm$0.7 & 28$\pm$6  \\
            & & & & & 36$\pm$3 & 739$\pm$26 & 1528$\pm$34 & 8019$\pm$175 & 6150$\pm$109 & 1583$\pm$3  \\
\hline 
L1582A-IRS5 & 05:31:50 +12:31:26.1 &    2.3$\pm$0.8 & 300$\pm$97 & I & 21$\pm$2 & 46$\pm$4 & 47$\pm$4 & 30$\pm$1 & 25$\pm$1 & 18$\pm$1  \\
                           & & & & & 16$\pm$1 & 11$\pm$1 & 17$\pm$2 & 8$\pm$1 & 18$\pm$2 & 16$\pm$4  \\
         & & & & & 19$\pm$2 & 1606$\pm$54 & 2596$\pm$63 & 18640$\pm$3771 & 15641$\pm$247 & 111$\pm$3  \\
\hline 
L1582A-IRS6 & 05:31:49 +12:31:58.3 &    4.4$\pm$1.3 & 453$\pm$122 & Flat & 74$\pm$6 & 104$\pm$8 & 115$\pm$9 & 123$\pm$1 & 156$\pm$2 & 106$\pm$1  \\
                     & & & & & 121$\pm$1 & 94$\pm$1 & 89$\pm$2 & 170$\pm$1 & 152$\pm$2 & 140$\pm$31  \\
        & & & & & 171$\pm$6 & 2473$\pm$66 & 5718$\pm$91 & 25461$\pm$441 & 26257$\pm$320 & 119$\pm$3  \\
\hline 
L1582A-IRS7 & 05:31:54+12:30:42.5 &    1.6$\pm$1.4 &  45$\pm$39 & 0 & 0.06$\pm$0.04 &  \nodata  & 0.13$\pm$0.06 & 0.76$\pm$0.17 & 0.11$\pm$0.03 & 0.66$\pm$0.08  \\
                             & & & & & 0.04$\pm$0.02 & 0.60$\pm$0.08 & 19$\pm$1 & 1.5$\pm$0.1 & 20$\pm$2 & 28$\pm$6  \\
          & & & & & 53$\pm$3 & 2113$\pm$61 & 3346$\pm$71 & 12882$\pm$314 & 7804$\pm$174 & 259$\pm$5  \\
 \hline
L1041-2-IRS1 & 20:37:19 +57:49:07.6 &    0.6$\pm$0.5 &  27$\pm$21 & 0 & \nodata  & 0.08$\pm$0.05 & 0.03$\pm$0.06 & 0.07$\pm$0.01 & 0.05$\pm$0.02 & 0.11$\pm$0.01  \\
                                & & & & & 0.06$\pm$0.02 & 0.04$\pm$0.02 & 0.16$\pm$0.03 & 0.22$\pm$0.04 & 0.39$\pm$0.04 & 0.87$\pm$0.30  \\
                              & & & & & 0.72$\pm$0.25 &  \nodata  &  \nodata  & 1432$\pm$88 & 2368$\pm$82 & 112$\pm$3  \\
 \hline
L1041-2-IRS2 & 20:37:12 +57:47:49.8 &    0.3$\pm$0.1 & 220$\pm$109 & I & 0.14$\pm$0.01 & 0.09$\pm$0.01 & 1.2$\pm$0.1 & 3.9$\pm$0.1 & 6.5$\pm$0.2 & 8.4$\pm$0.1  \\
                          & & & & & 9.8$\pm$0.3 & 14$\pm$0.3 & 13$\pm$1 & 17$\pm$1 & 19$\pm$1 & 58$\pm$21  \\
                 & & & & & 54$\pm$3 & 47$\pm$10 & 72$\pm$12 & 538$\pm$64 & 1421$\pm$74 & 679$\pm$8  
\enddata
\tablenotetext{a}{The uncertainties of $L_{bol}$ were estimated by propagating the errors of distance and flux densities.}
\tablenotetext{b}{The uncertainties of $T_{bol}$ were calculated by propagating the errors of flux densities.}
\tablenotetext{c}{The uncertainties of flux densities are the errors resulted in performing the aperture photometry for the sources.}
\label{tbl:flux}
\end{deluxetable}

\begin{deluxetable}{ccccccccc}
\tabletypesize{\scriptsize}
\tablewidth{0pt}
\tablecaption{Possible heating sources near the dense cores}
\tablehead{Core & Source & Position & D\tablenotemark{a} & Spt.\tablenotemark{b} & R$_{\ast}$\tablenotemark{c} & T$\rm_{bol}$\tablenotemark{c} & L$\rm_{bol}$\tablenotemark{d} & D$\rm_{\ast-cl}$\tablenotemark{e} \\
& & (J2000) & (pc) & & (pc) & (K) & (L$_{\odot}$) & (pc) }
\startdata
CB22 & HD 282383 & 04:41:02+29:43:41 & 139 & F8 & 1.2 & 6491 & 2.3 & 0.4 \\
CB22 & HD 29537 & 04:40:22+29:58:20 & 87 & F0 & 1.5 & 7706 & 7.1 & 0.2 \\
L1517B & HD 31293 & 04:55:45+30:33:04 & 139 & A0 & 2.4 & 10252 & 57.0 & 0.3 \\
L1512 & HR 1627 & 05:04:36+32:19:13 & 97 & A4 & 2.1 & 8507 & 20.7 & 0.7 \\
L1234 & IRAS 23158+6208 & 23:17:58+62:24:28 & 300 & YSO & \nodata & 2027 & 47.1 & 0.2 
\enddata
\tablenotetext{a}{The distance of the source was derived using the Hipparcos parallax measurements obtained from \citet{vanLeeuwen:2007dc}. The distance of IRAS 23158+6208 was assumed to be the same as that of the parent core.}
\tablenotetext{b}{The spectral types of the stars are from \citet{Nesterov:1995vx}.}
\tablenotetext{c}{The radius and bolometric temperature of the stars were obtained from \citet{Cox:2000ua}. For IRAS 23158+6208, the bolometric temperature was estimated from its spectral energy distribution.}
\tablenotetext{d}{The bolometric luminosity was derived from the bolometric temperature and the radius of the star. For IRAS 23158+6208, the bolometric luminosity was estimated from the spectral energy distribution of the source.}
\tablenotetext{e}{The projected distance between the star and the core was calculated by assuming that the star and the core have the same distance from us.}
\label{tbl:origin}
\end{deluxetable}

\begin{deluxetable}{cccccccccc}
\tabletypesize{\scriptsize}
\tablewidth{0pt}
\tablecaption{Observed and Virial masses, and their related quantities of the six starless cores}
\tablehead{Object & $\rm V_{LSR}(N_{2}H^{+})$ & $\Delta \rm V(N_{2}H^{+})$ & $\rm T_{ex}$ & R$\rm_{obs}$ & N$\rm (H_2)_{peak}$ & M$\rm_{obs}$ & M$\rm_{A_V}$ & M$\rm_{vir}$ & $\rm \alpha_{vir}$ \\
& ($\rm km~s^{-1}$) & ($\rm km~s^{-1}$) & (K) & (pc) & ($\rm \times10^{21}~cm^{-2}$) & (M$_{\sun}$) & (M$_{\sun}$) & (M$_{\sun}$) & \\
(1) & (2) & (3) & (4) & (5) & (6) & (7) & (8) & (9) & (10)}
\startdata
CB22N    & 5.96$\pm$0.01 & 0.23$\pm$0.02 & 16$\pm$2 & 0.04 & 4.3$\pm$3.1 & 0.52$\pm$0.38 & 0.33$^{+0.14}_{-0.06}$ & 1.86$^{+0.32}_{-0.81}$ & 3.6$\pm$3.0 (5.7$\pm$3.5) \\
CB22S    & 5.96$\pm$0.01 & 0.21$\pm$0.02 & 24$\pm$3 & 0.03 & 5.2$\pm$3.7 & 0.41$\pm$0.30 & 0.11$^{+0.05}_{-0.02}$ & 1.42$^{+0.23}_{-0.61}$ & 3.5$\pm$3.0 (12.6$\pm$7.7) \\
L1517B  & 5.81$\pm$0.01 & 0.21$\pm$0.02 & 6$\pm$1 & 0.04 & 7.4$\pm$5.3 & 0.72$\pm$0.52 & 0.76$^{+0.33}_{-0.12}$ & 2.08$^{+0.36}_{-0.91}$ & 2.9$\pm$2.5 (2.7$\pm$1.7) \\
L1512   & 7.07$\pm$0.01 & 0.19$\pm$0.01 & 15$\pm$1 & 0.04 & 8.9$\pm$6.3 & 0.76$\pm$0.56 & 0.28$^{+0.12}_{-0.05}$ & 1.66$^{+0.26}_{-0.71}$ & 2.2$\pm$1.8 (6.0$\pm$3.7) \\
L1621-1 & 1.78$\pm$0.01 & 0.15$\pm$0.01 & 12$\pm$1 & 0.07 & 2.8$\pm$2.0  & 0.98$\pm$0.76 & 1.26$^{+0.64}_{-0.39}$ & 3.03$^{+0.92}_{-1.52}$ & 3.1$\pm$2.9 (2.4$\pm$1.7) \\
L1234   & -5.05$\pm$0.02 & 0.32$\pm$0.05 & 4$\pm$1 & 0.04 & 3.8$\pm$2.7  & 0.33$\pm$0.24 & 0.89$^{+0.40}_{-0.19}$ & 2.56$^{+0.59}_{-1.18}$ & 7.8$\pm$6.8 (2.9$\pm$1.9) \\
CB246-2 & -0.82$\pm$0.01 & 0.26$\pm$0.02 & 5$\pm$1 & 0.04 & 6.0$\pm$4.3 & 0.56$\pm$0.41 & 0.53$^{+0.23}_{-0.08}$ & 2.06$^{+0.34}_{-0.89}$ & 3.7$\pm$3.1 (3.9$\pm$2.3) \\
\enddata
\tablecomments{Col. (1): The source name. Col. (2): The average LSR velocity (derived from seven hyperfine Gaussian fit to the N$_2$H$^+$ lines) of the spectra within the Half Maximum (HM) contour of the N$_2$H$^+$ intensity map. Col. (3): The average line width (derived from seven hyperfine Gaussian fit to the N$_2$H$^+$ lines) of the spectra within the HM contour of the N$_2$H$^+$ intensity map. Col. (4): The average excitation temperature (derived from seven hyperfine Gaussian fit to the N$_2$H$^+$ lines) 
for the spectra within the HM contour of the N$_2$H$^+$ intensity map. Col. (5): The effective radius of an area enclosing the HM contour of the N$_2$H$^+$ intensity map. 
Col (6): The column density of H$_2$ toward the $\rm N_2H^+$ (1-0) intensity peak position. Col. (7): The H$_2$ mass (derived from N$_2$H$^+$ (1-0) data) of the core within the HM contour of its N$_2$H$^+$ intensity map. Col. (8): The H$_2$ mass (derived from the 2MASS extinction data) of the core within the HM contour of its N$_2$H$^+$ intensity map. Col. (9): The virial mass for the core within the HM contour of its N$_2$H$^+$ intensity map. Col. (10): The virial parameter $\alpha = M_{vir}/M_{obs}$. }
\label{tbl:virial}
\end{deluxetable}

\clearpage
\renewcommand{\thefigure}{1-{\alph{figure}}}

\begin{figure}
\includegraphics[bb=0 0 761 567,scale=.6]{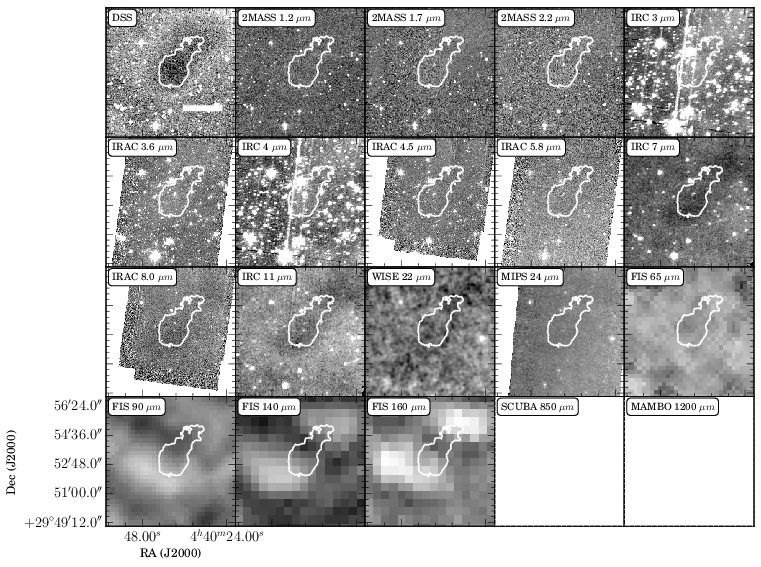}
\caption{The multi-wavelength images of CB22 ranging from optical R-band to 1200 {\micron} wavelengths. These images were obtained from the DSS, 2MASS, \textit{Spitzer}, \textit{WISE}, \textit{AKARI}, \textit{JCMT}, and \textit{IRAM}. The contour indicates the approximate boundary of the dense core, which is a half minimum contour level of an extinction region in the optical image. The horizontal bar on the DSS image is a scale of 0.1 pc. The empty panel indicates non-availability of continuum data. \label{fig:multiCB22}}
\end{figure}

\begin{figure}
\includegraphics[bb=0 0 761 567,scale=.6]{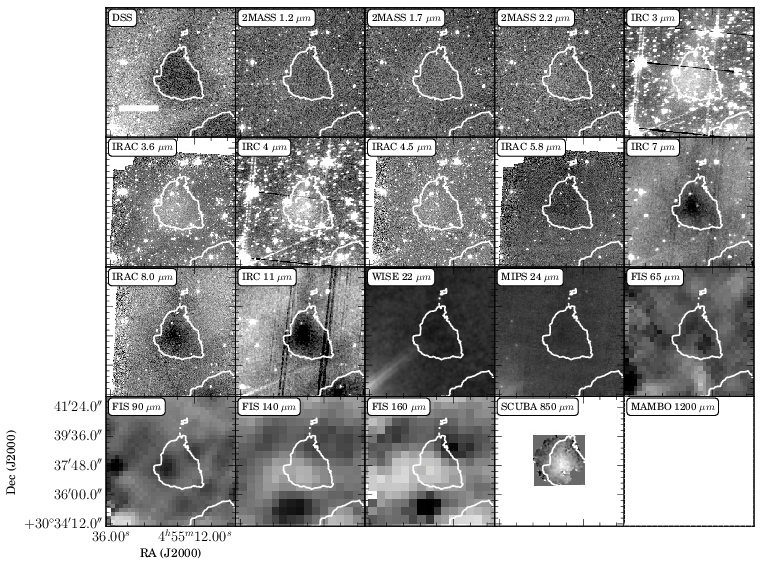}
\caption{Same as Figure \ref{fig:multiCB22} but of L1517B.\label{fig:multiL1517B}}
\end{figure}

\begin{figure}
\includegraphics[bb=0 0 761 567,scale=.6]{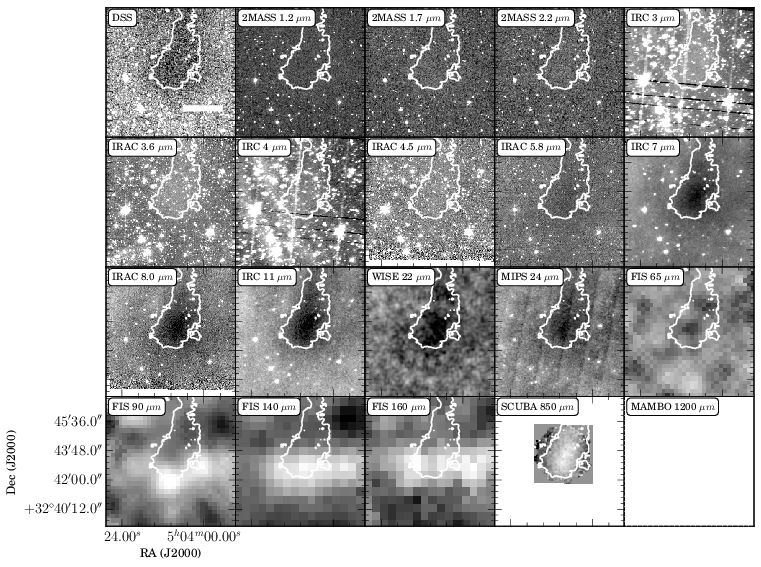}
\caption{Same as Figure \ref{fig:multiCB22} but of L1512.\label{fig:multiL1512}}
\end{figure}

\begin{figure}
\includegraphics[bb=0 0 761 567,scale=.6]{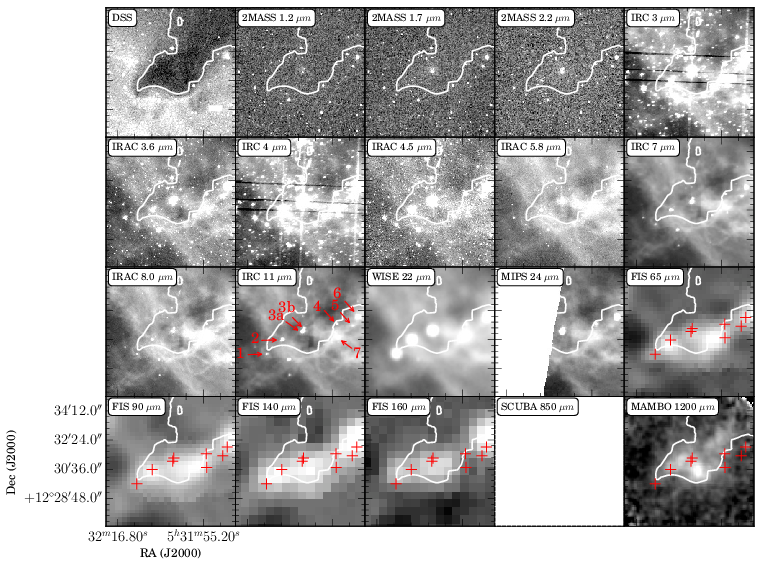}
\caption{Same as Figure \ref{fig:multiCB22} but of L1582A. The crosses and the arrows with numbers indicate positions of the embedded YSOs.\label{fig:multiL1582A}}
\end{figure}

\begin{figure}
\includegraphics[bb=0 0 761 567,scale=.6]{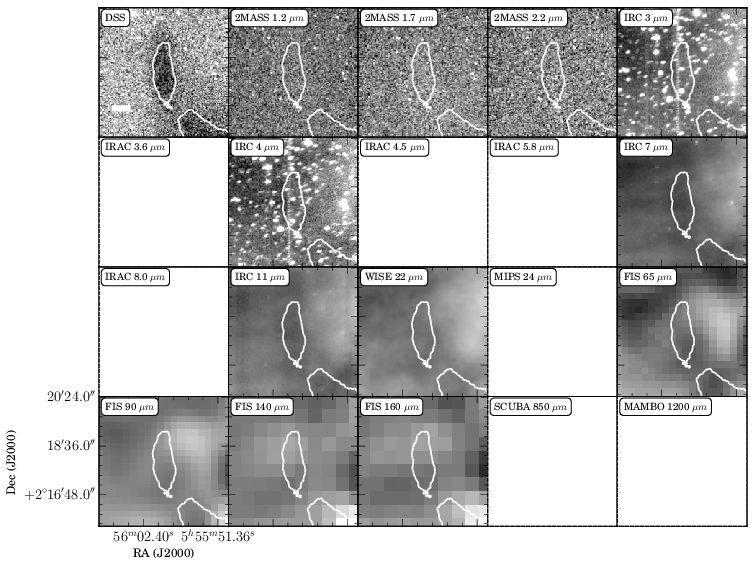}
\caption{Same as Figure \ref{fig:multiCB22} but of L1621-1. \label{fig:multiL1621-1}}
\end{figure}

\begin{figure}
\includegraphics[bb=0 0 761 567,scale=.6]{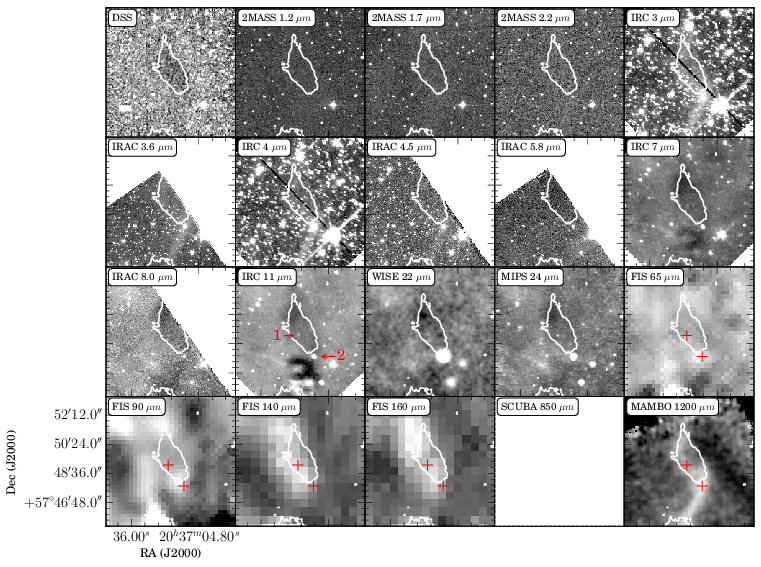}
\caption{Same as Figure \ref{fig:multiL1582A} but of L1041-2. \label{fig:multiL1041-2}}
\end{figure}

\begin{figure}
\includegraphics[bb=0 0 761 567,scale=.6]{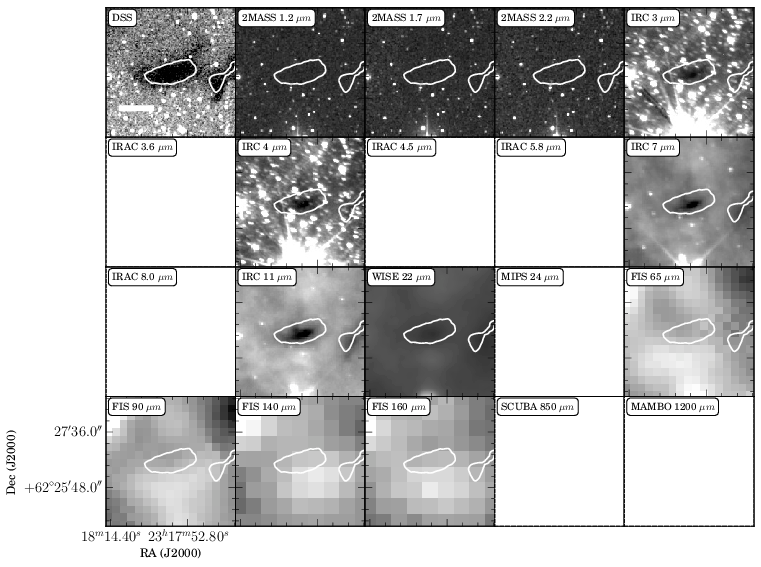}
\caption{Same as Figure \ref{fig:multiCB22} but of L1234.\label{fig:multiL1234}}
\end{figure}

\begin{figure}
\includegraphics[bb=0 0 761 567,scale=.6]{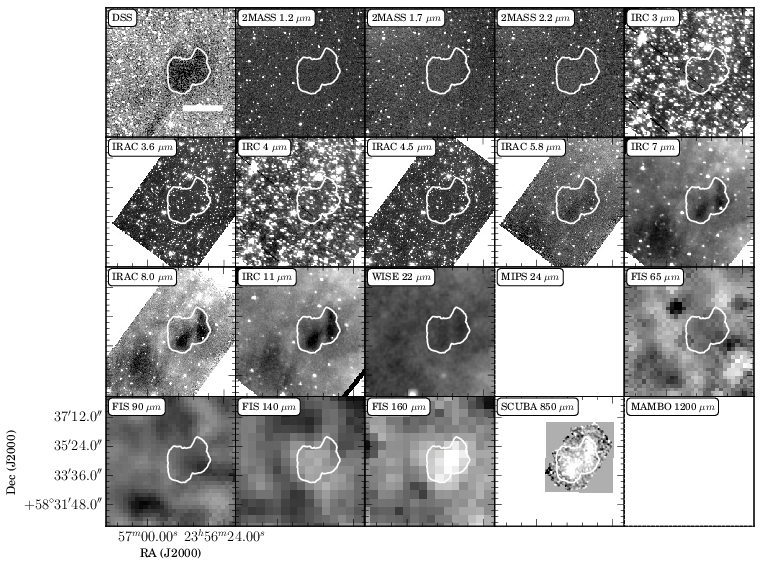}
\caption{Same as Figure \ref{fig:multiCB22} but of CB246-2. \label{fig:multiCB246-2}}
\end{figure}

\setcounter{figure}{1} \renewcommand{\thefigure}{\arabic{figure}}

\begin{figure}
$\begin{array}{ccc}
\includegraphics[bb=0 0 550 330,scale=0.28]{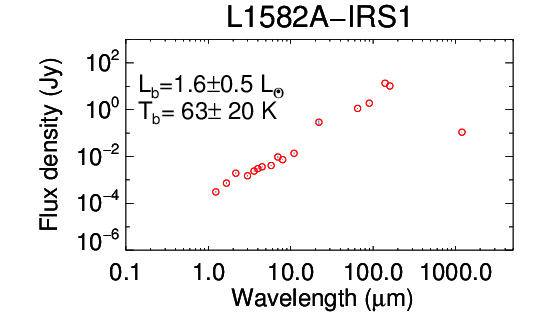}&
\includegraphics[bb=0 0 550 330,scale=0.28]{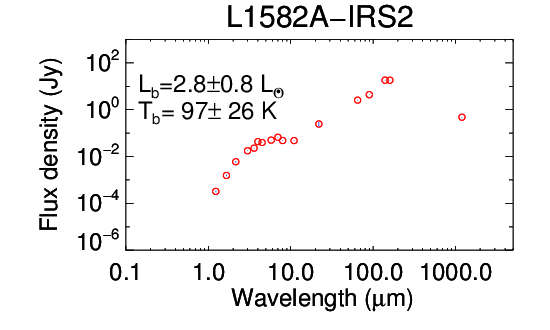}&
\includegraphics[bb=0 0 550 330,scale=0.28]{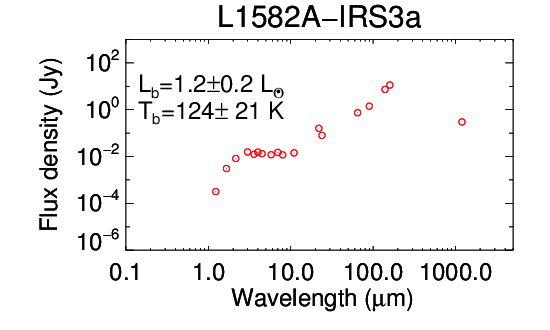}\\
\includegraphics[bb=0 0 550 330,scale=0.28]{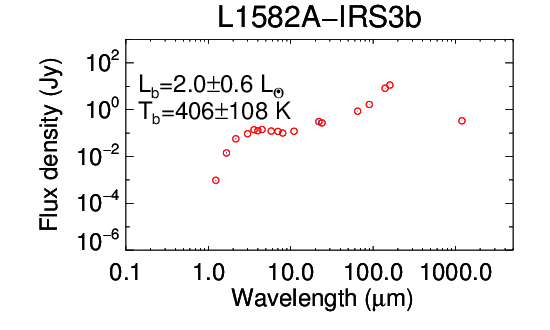}&
\includegraphics[bb=0 0 550 330,scale=0.28]{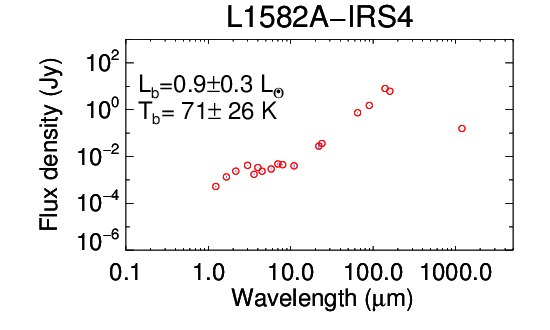}&
\includegraphics[bb=0 0 550 330,scale=0.28]{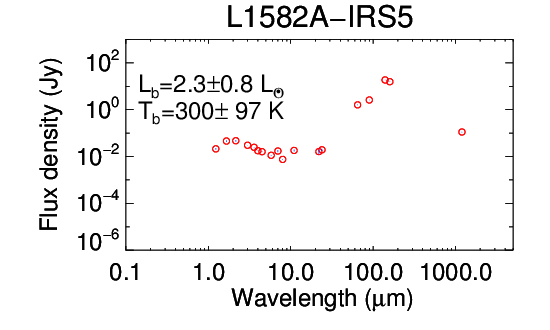}\\
\includegraphics[bb=0 0 550 330,scale=0.28]{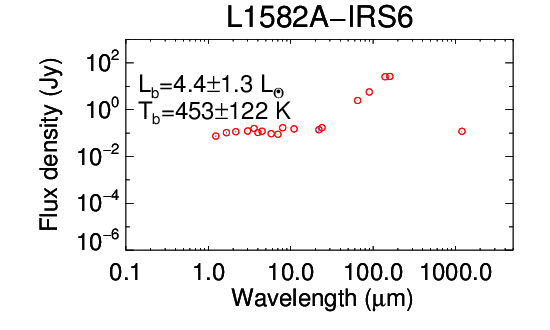}&
\includegraphics[bb=0 0 550 330,scale=0.28]{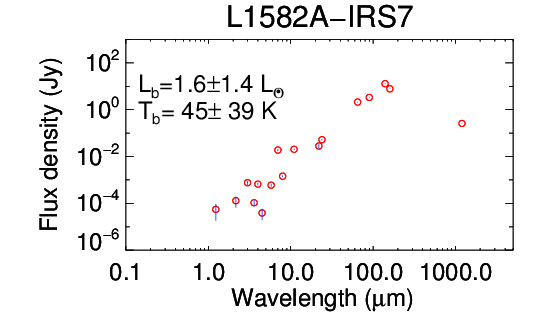}&
\includegraphics[bb=0 0 550 330,scale=0.28]{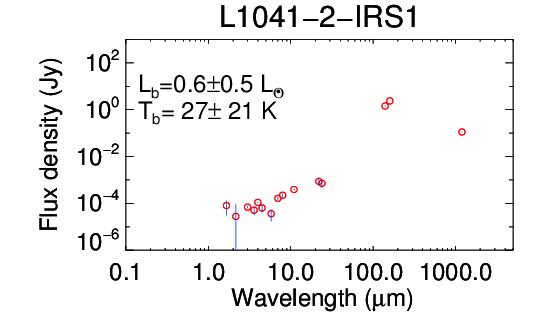}\\
\includegraphics[bb=0 0 550 330,scale=0.28]{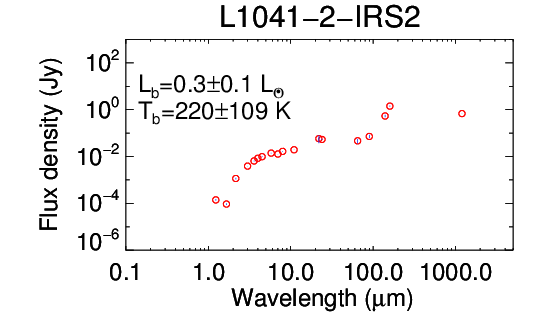} 
\end{array}$
\caption{The spectral energy distributions for possible YSOs found in dense cores L1582A and L1041-2. In the bolometric luminosity calculation, we assumed that the distances of protostars are the same as those of the parent cores (400 pc for L1582A and 440 pc for L1041-2). 
\label{fig:sed}}
\end{figure}

\begin{figure}
\center
$\begin{array}{cc}
\includegraphics[bb=0 0 1376 2964,scale=.15]{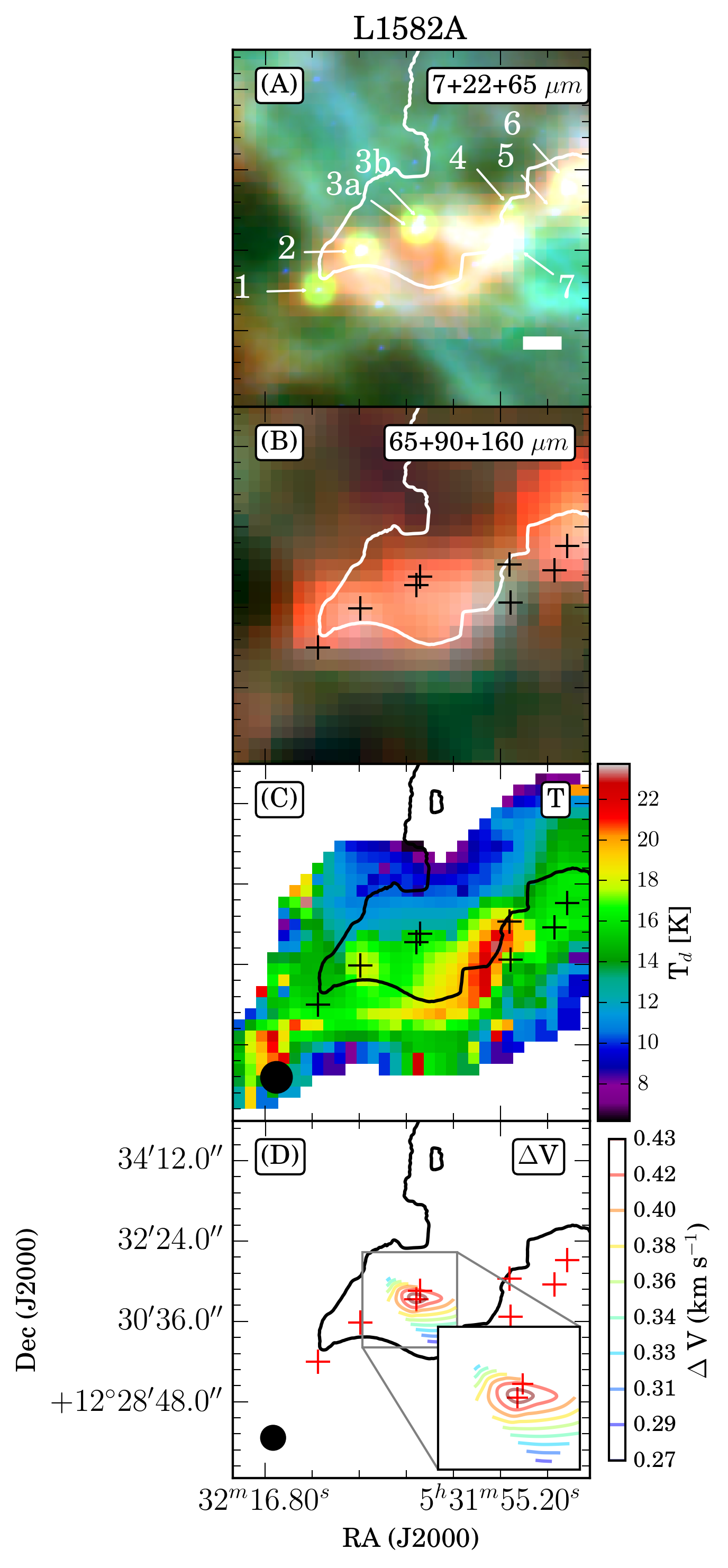} &
\includegraphics[bb=0 0 1376 2964,scale=.15]{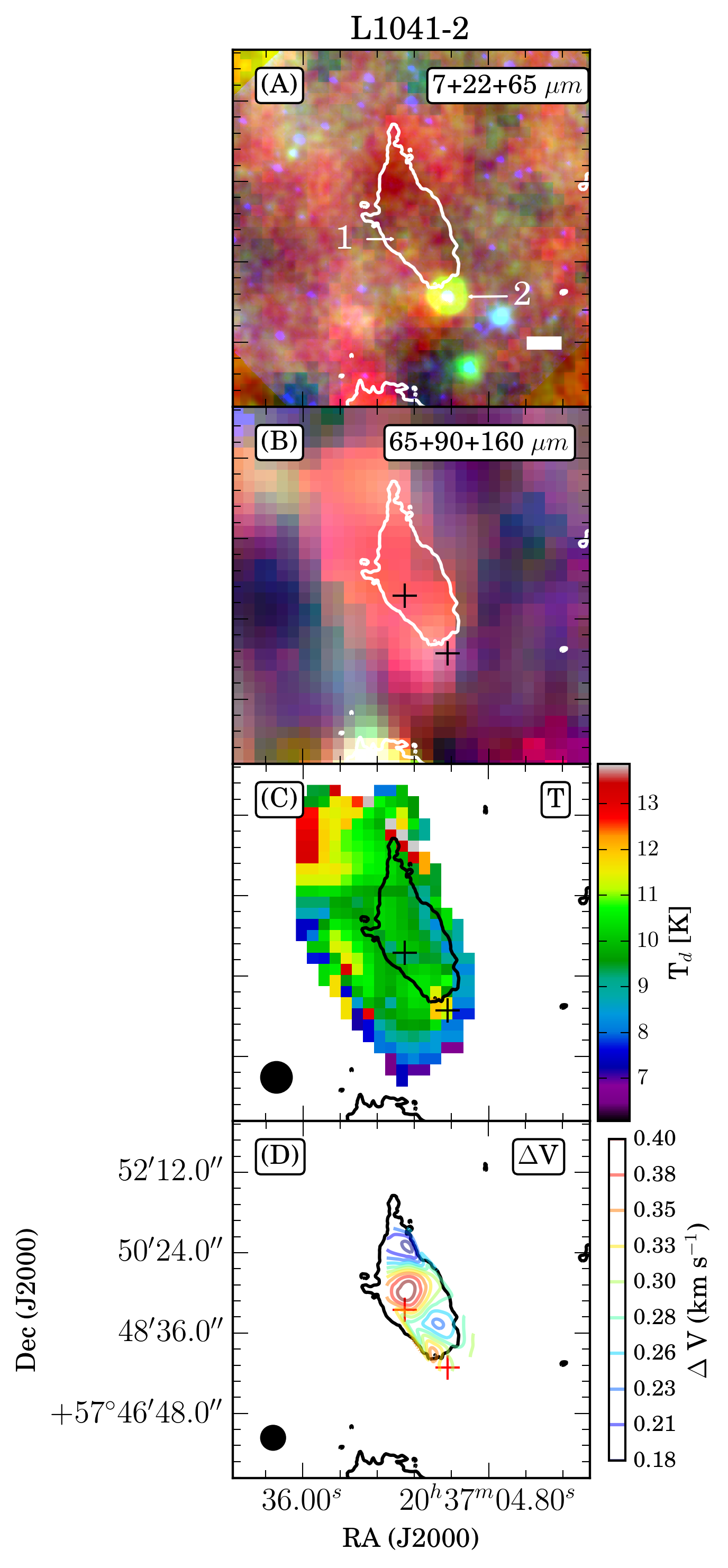} 
\end{array}$
\caption{Color-composite, temperature, and spectral line width maps for two protostellar cores. The (A) and (B) panels are color-composite images produced using 7 {\micron} (Blue) + 22 {\micron} (Green) + 65 {\micron} (Red) and 65 {\micron} (Blue) + 90 {\micron} (Green) + 160 {\micron} (Red), respectively. The (C) panels are the maps for dust temperature distributions. The (D) panels are the contour maps of line widths derived from the N$_2$H$^+$ line profiles. The thick white/black contours mark the approximate boundary of the dense core as explained in Figure \ref{fig:multiCB22}. The white arrows with numbers and the crosses mark the positions of the embedded YSOs. The white horizontal bars on the (A) panels are the scale of 0.1 pc. The circles on (C) and (D) panels indicate the beam sizes of 41{\arcsec} and 32{\arcsec}, respectively. 
\label{fig:rgb1}}
\end{figure}

\begin{figure}
\center
$\begin{array}{c}
\includegraphics[bb=0 0 1551 800,scale=.25]{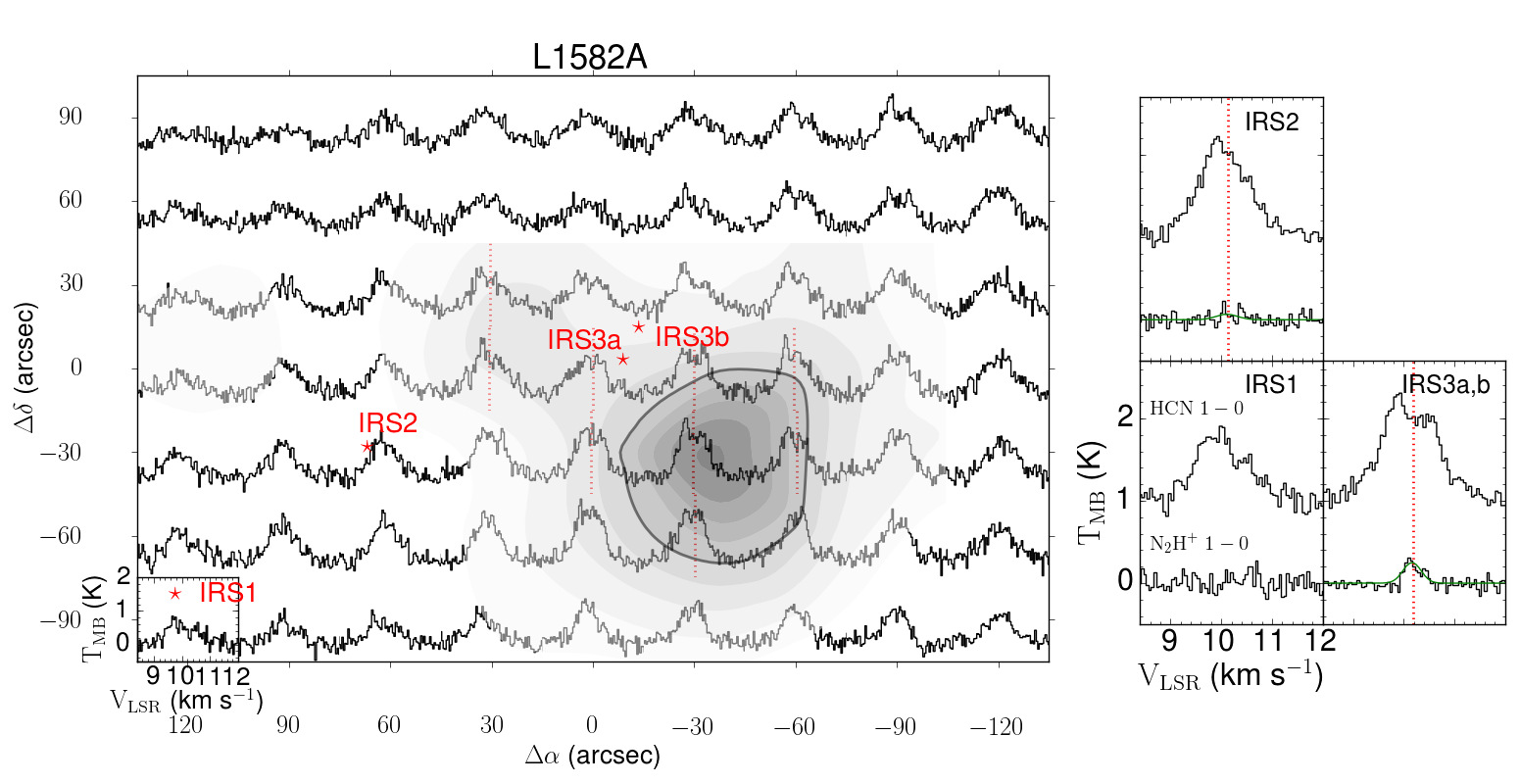} \\
\includegraphics[bb=0 0 1046 800,scale=.25]{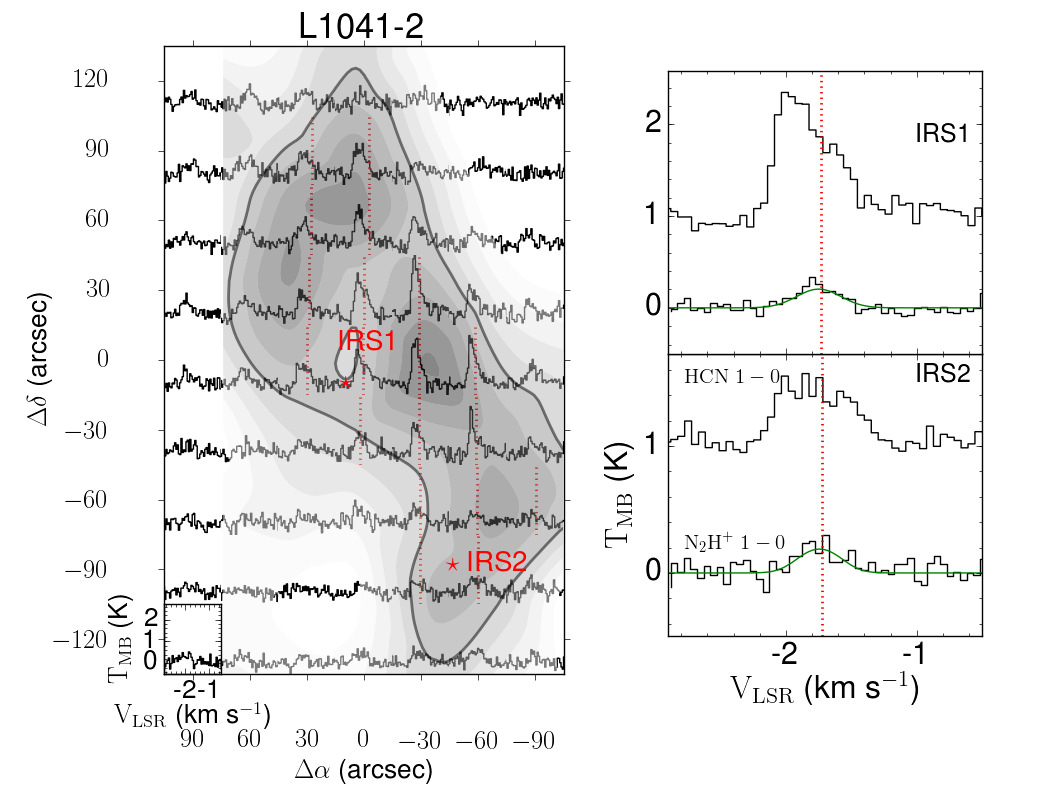} 
\end{array}$
\caption{The HCN (1-0) and N$_2$H$^+$ (1-0) line profile maps of two protostellar cores. The left panel shows the line profile maps of the main component (F=2-1; 88631.8473 MHz) of the HCN (1-0) lines superposed with the half maximum contour of integrated intensity of N$_2$H$^+$. The minimum level, maximum level, and interval of the integrated intensity of N$_2$H$^+$ are as follows; L1582A: 0.27, 2.66, 1.33 K km s$^{-1}$, L1041-2: 0.13, 1.26, 0.63 K km s$^{-1}$. The right panel displays the average spectra of HCN F=2-1 on the top and the isolated component (F$_1$F=01-12; 93176.2580 MHz) of the N$_2$H$^+$(1-0) line on the bottom over the embedded point sources. The green line is the Gaussian-fitted line of hyperfine structure. The dashed line in red on the profiles indicates the Gaussian fit velocities of N$_2$H$^+$(1-0) line. The symbol star in red is the position of YSO. The X and Y axes on the left panel indicate offsets in arc second from the given coordinates of each dense core. The X and Y axes of the inset put at left bottom corner of the left panel mark the LSR velocity in km s$^{-1}$ and the main beam temperature (T$\rm_{MB}$) in K.
\label{fig:HCNmap1}}
\end{figure}

\setcounter{figure}{0} 
\renewcommand{\thefigure}{5-{\alph{figure}}}

\begin{figure}
\center
$\begin{array}{c}
\includegraphics[bb=0 0 2729 939,scale=.15]{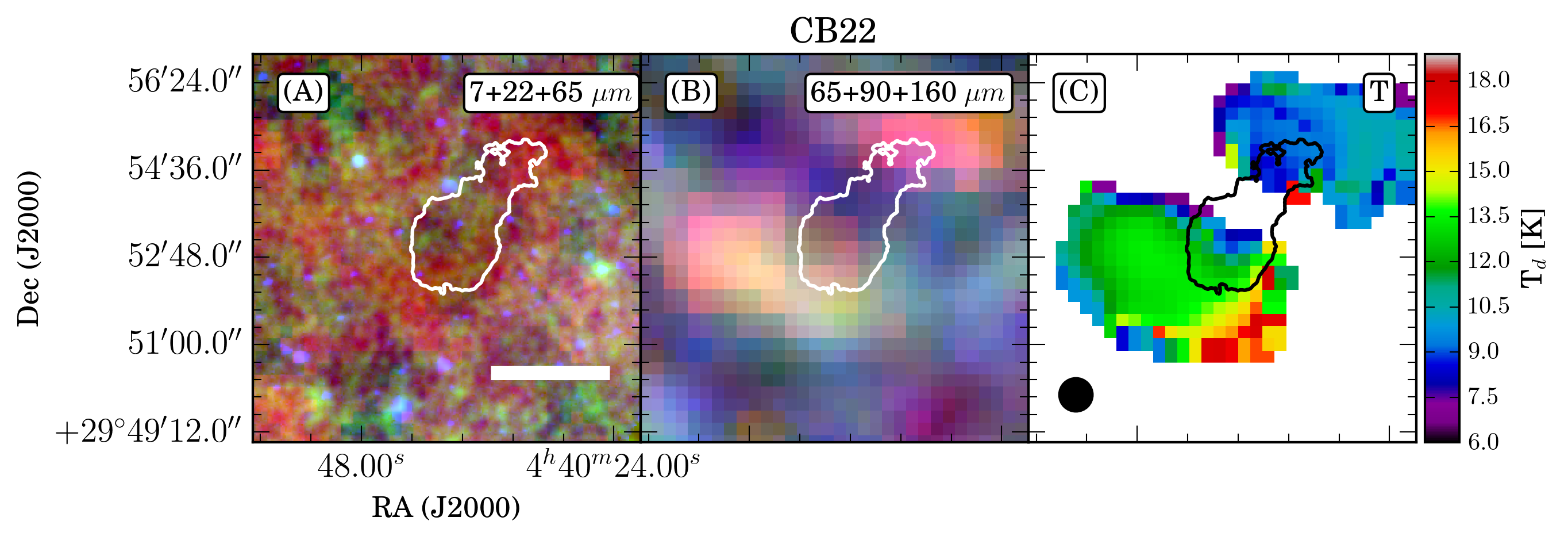} \\
\includegraphics[bb=0 0 2729 939,scale=.15]{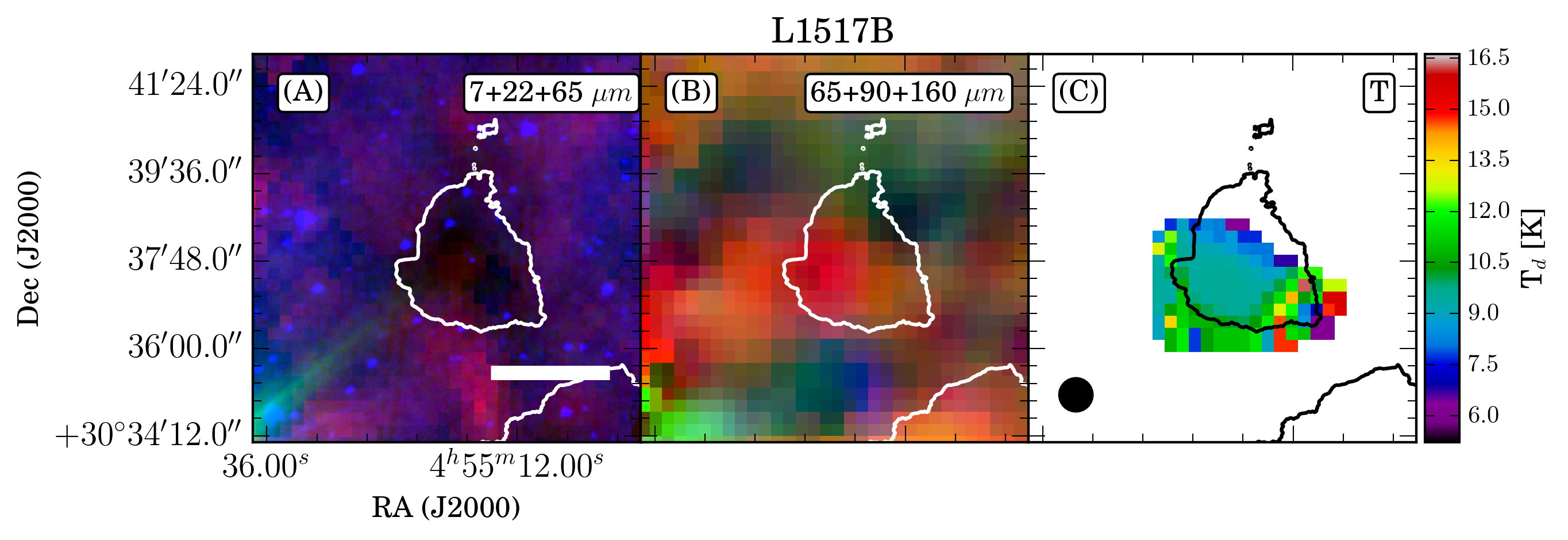} \\
\includegraphics[bb=0 0 2729 939,scale=.15]{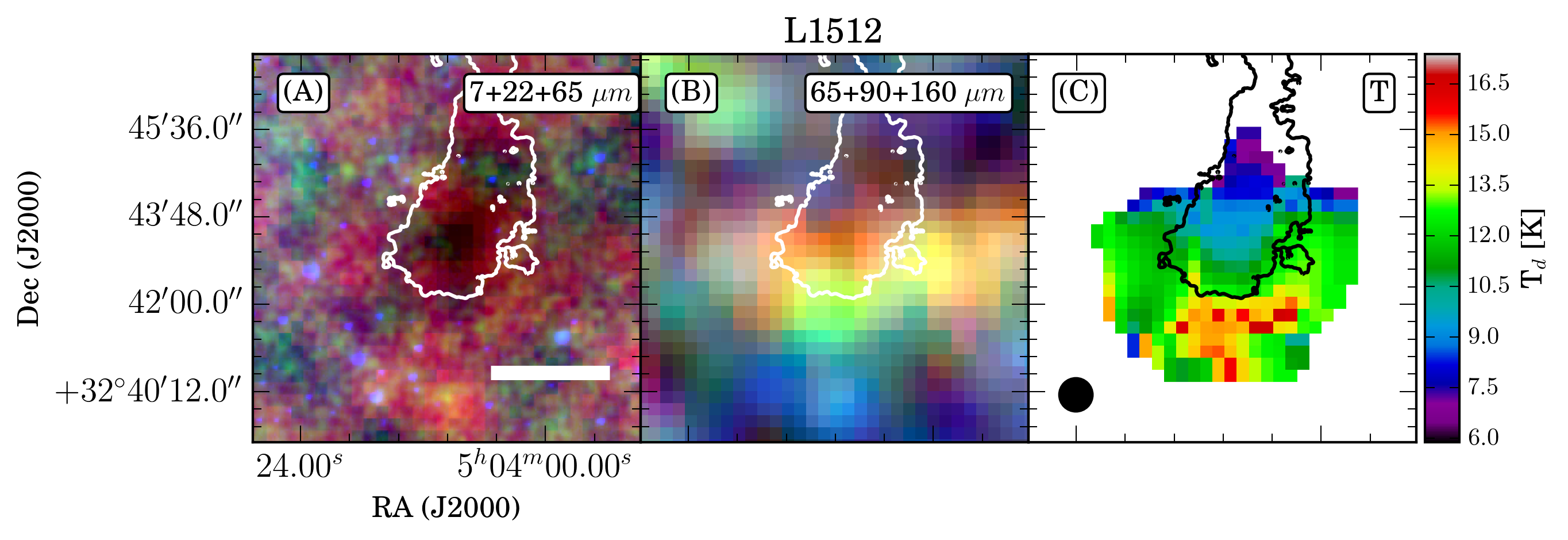} \\
\end{array}$
\caption{Same as Figure \ref{fig:rgb1} but the color-composite and temperature maps for three starless cores, CB22, L1517B, and L1512. \label{fig:rgb2}}
\end{figure}

\begin{figure}
\center
$\begin{array}{c}
\includegraphics[bb=0 0 2667 939,scale=.15]{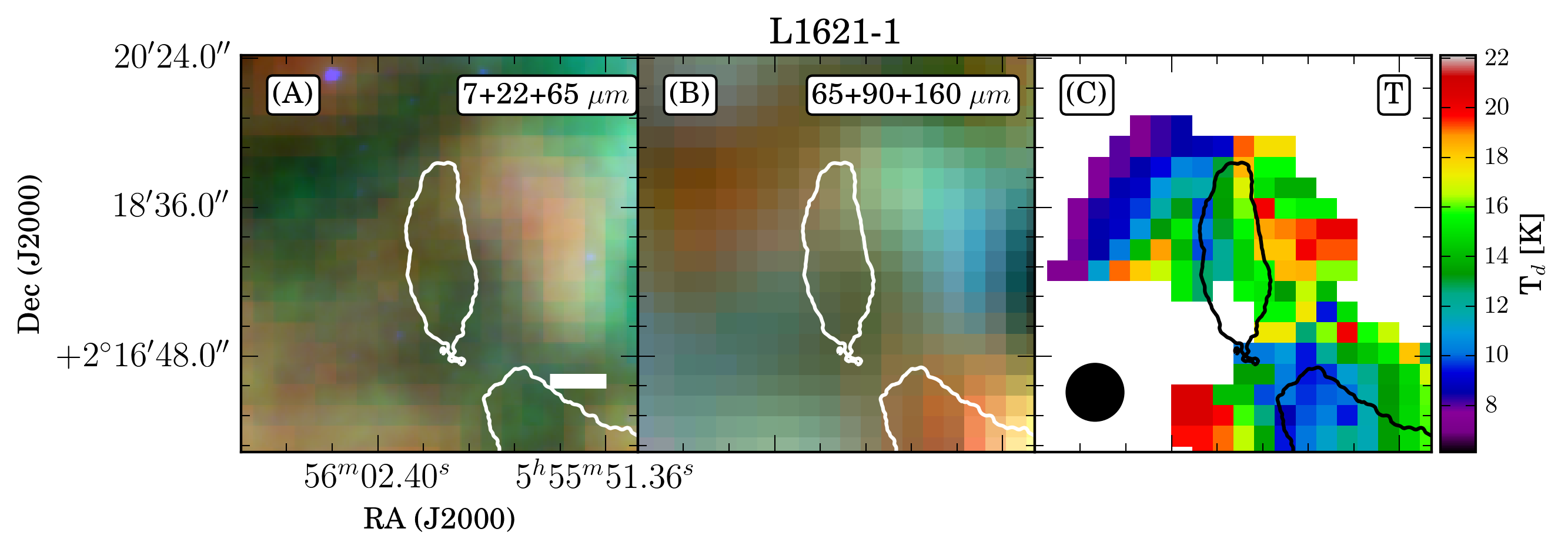} \\
\includegraphics[bb=0 0 2696 939,scale=.15]{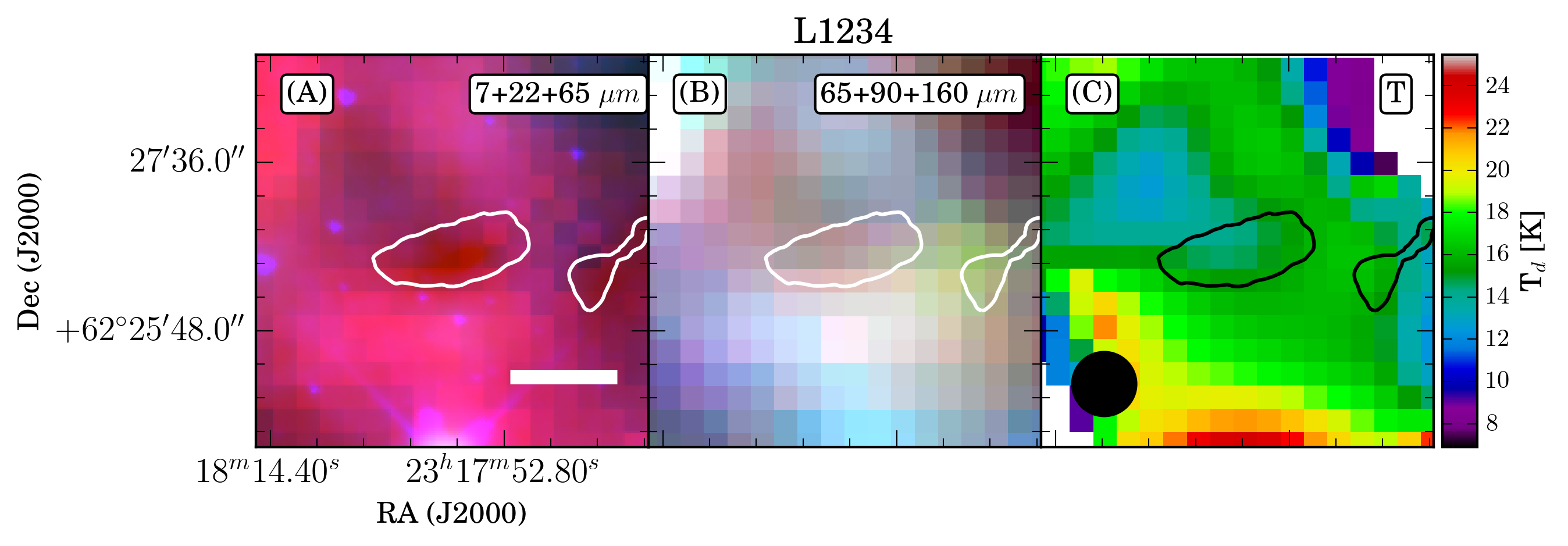} \\
\includegraphics[bb=0 0 2729 939,scale=.15]{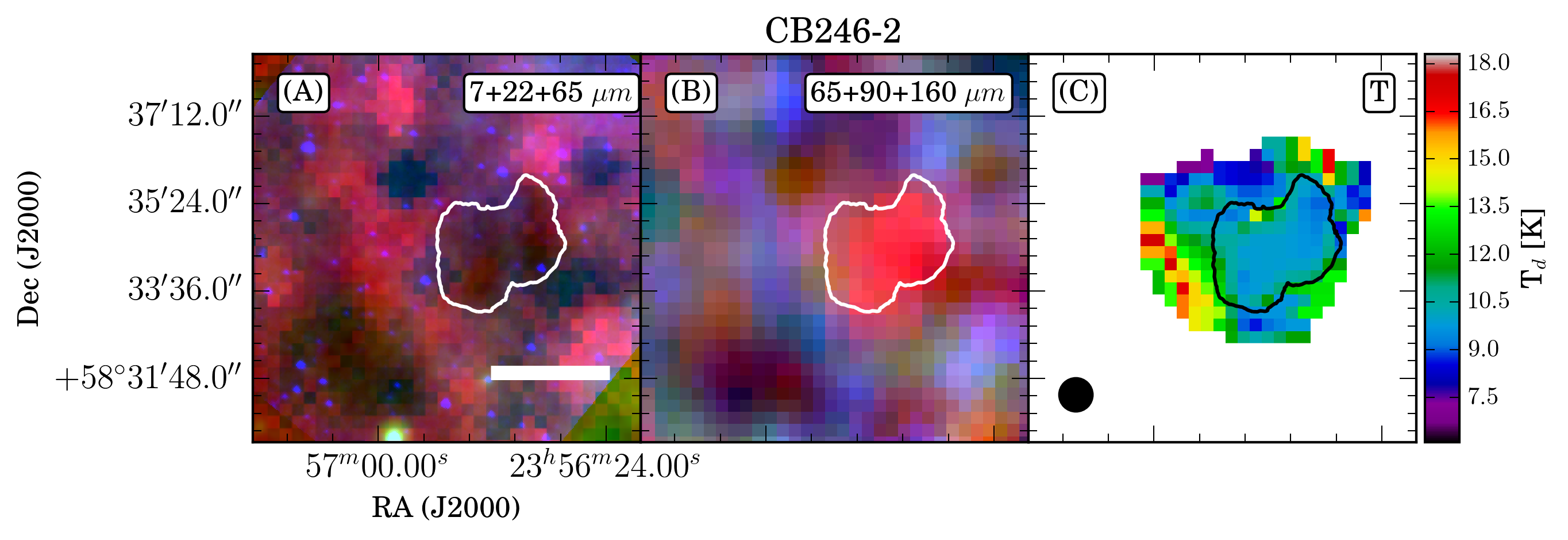} 
\end{array}$
\caption{Same as Figure \ref{fig:rgb1} but the color-composite and temperature maps for three starless cores, L1621-1, L1234, and CB246-2. \label{fig:rgb3}}
\end{figure}

\setcounter{figure}{5} 
\renewcommand{\thefigure}{\arabic{figure}}

\begin{figure}
\center
$\begin{array}{cc}
\includegraphics[bb=0 0 618 553,scale=.30]{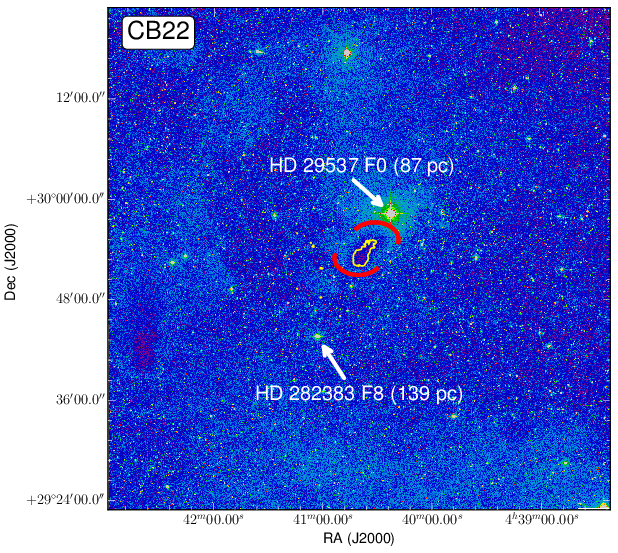} &
\includegraphics[bb=0 0 618 553,scale=.30]{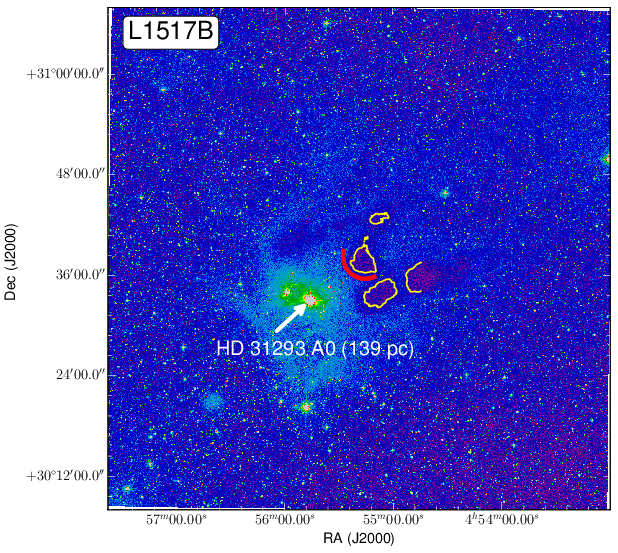} \\
\includegraphics[bb=0 0 633 553,scale=.30]{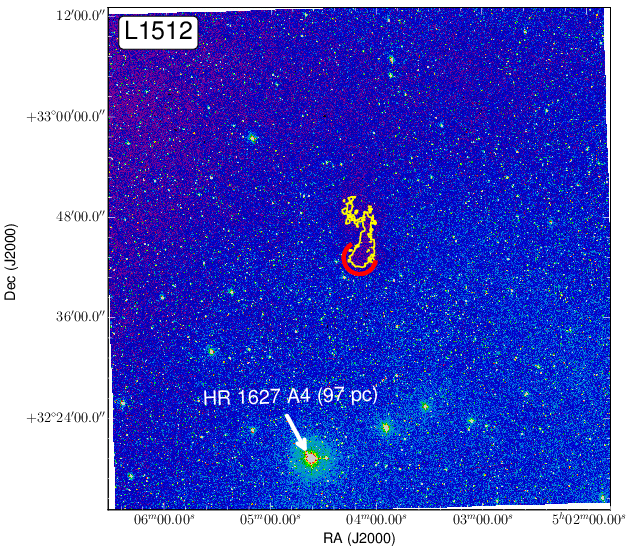} &
\includegraphics[bb=0 0 646 559,scale=.30]{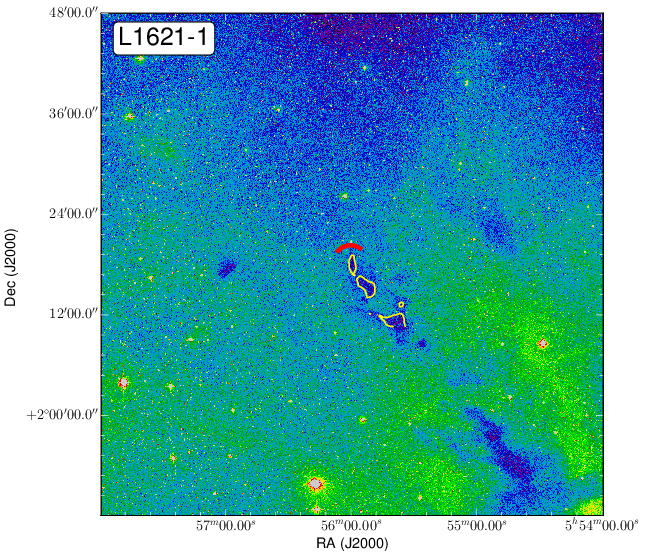} \\
\includegraphics[bb=0 0 641 553,scale=.30]{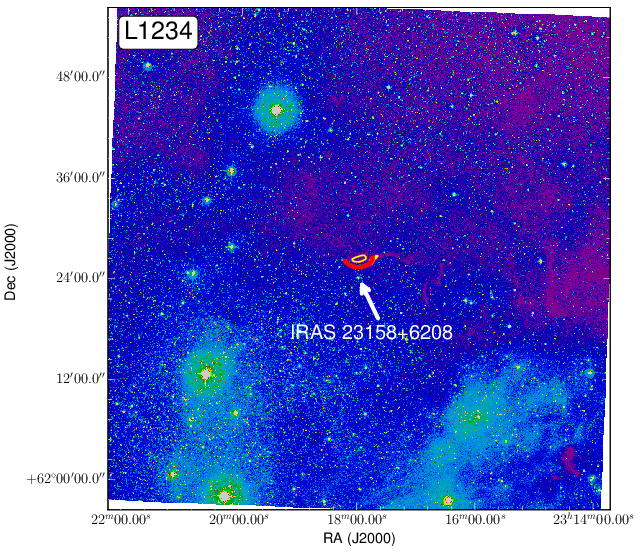} &
\includegraphics[bb=0 0 618 553,scale=.30]{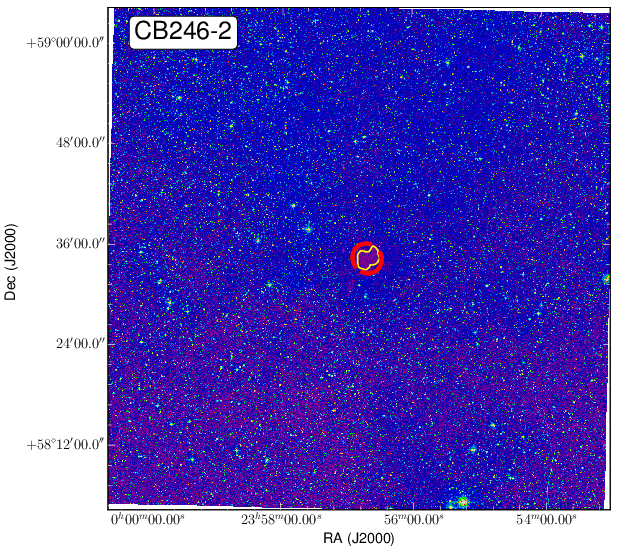} 
\end{array}$
\caption{DSS optical images of $1^{\circ}\times1^{\circ}$ for starless cores, CB22, L1517B, L1512, L1621-1, L1234, and CB246-2 and nearby heating sources. The yellow contour displays the approximate boundary of each dense core, which is the half minimum contour level of its extinction region in the optical image. The curves in red indicate the direction toward which the envelopes show a relative enhancement in temperature. The arrow in white is to indicate possible heating objects for which the spectral type and the distance are known from \citet{Nesterov:1995vx} and \citet{vanLeeuwen:2007dc}. \label{fig:dss_sw_w}}
\end{figure}

\setcounter{figure}{0} 
\renewcommand{\thefigure}{7-{\alph{figure}}}

\begin{figure}
\center
$\begin{array}{c}
\includegraphics[bb=0 0 1377 800,scale=.25]{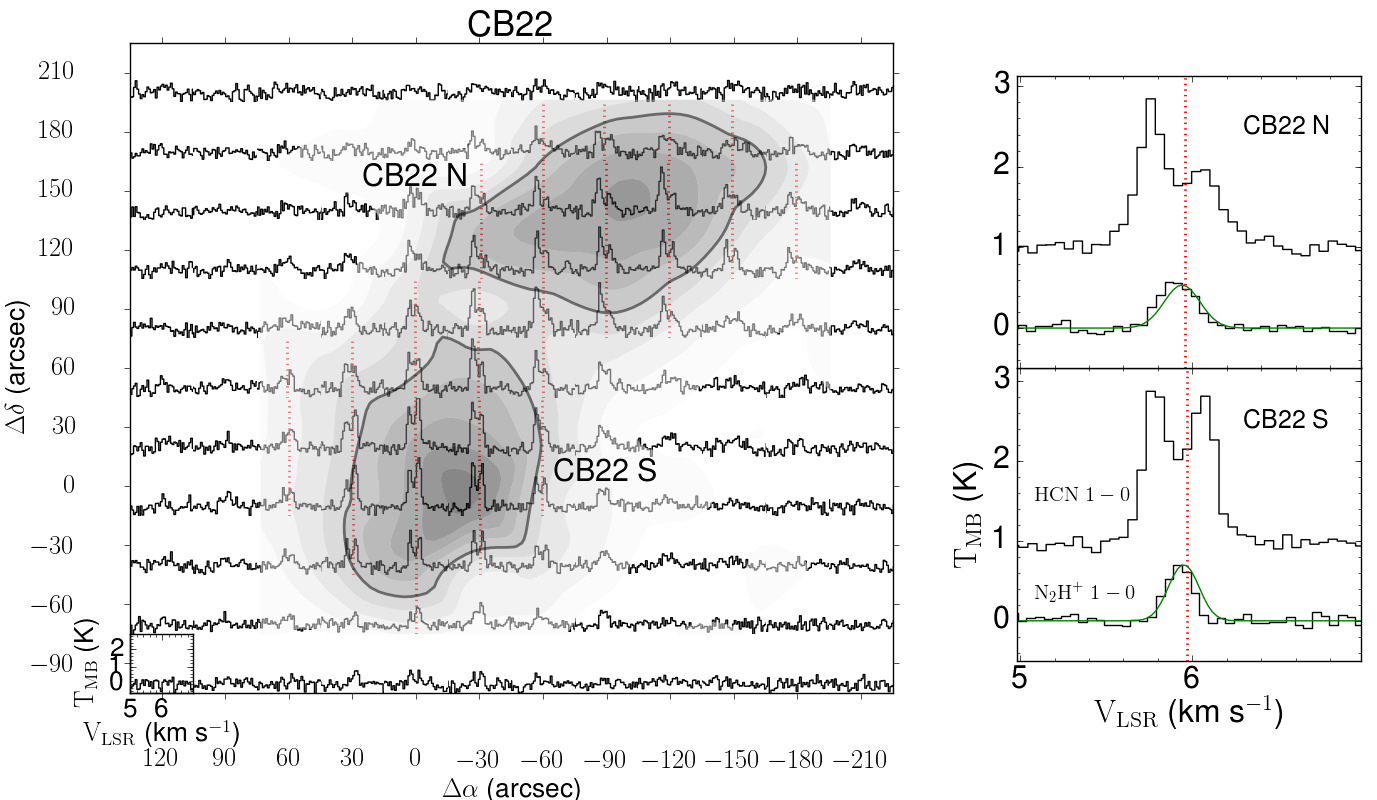} \\
\includegraphics[bb=0 0 1151 800,scale=.24]{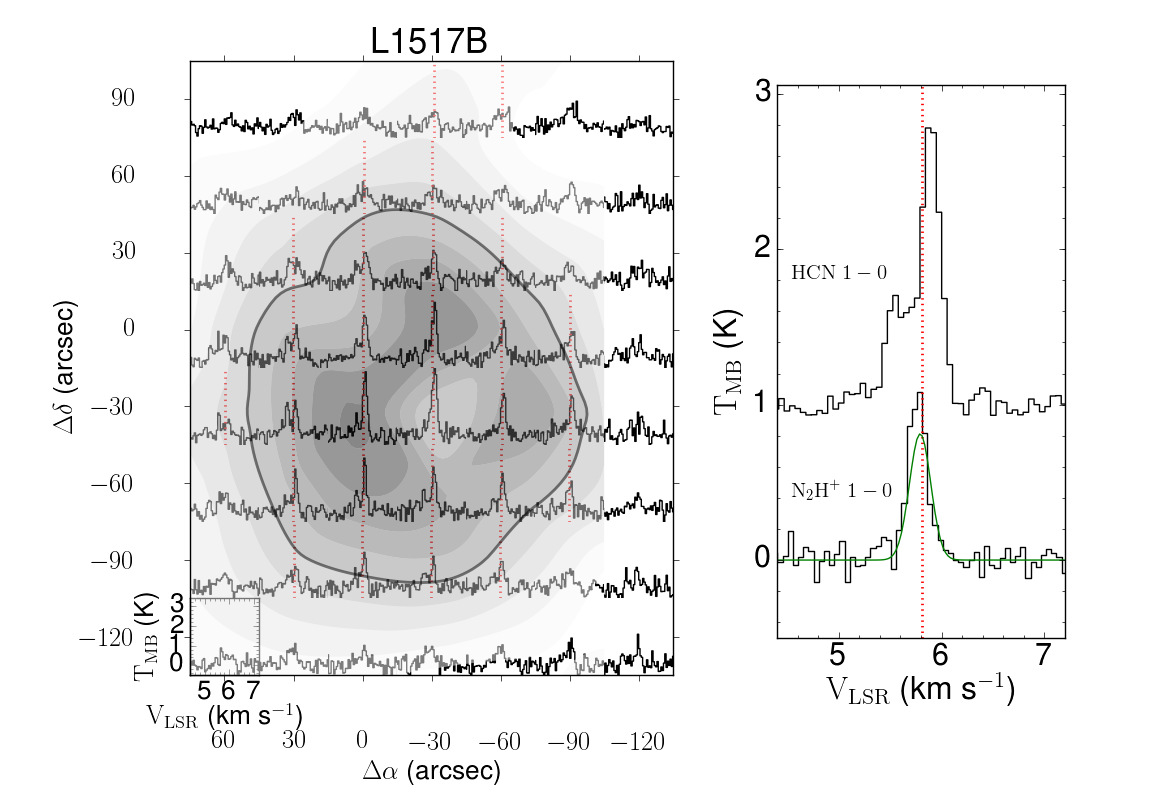} \\
\includegraphics[bb=0 0 1172 800,scale=.24]{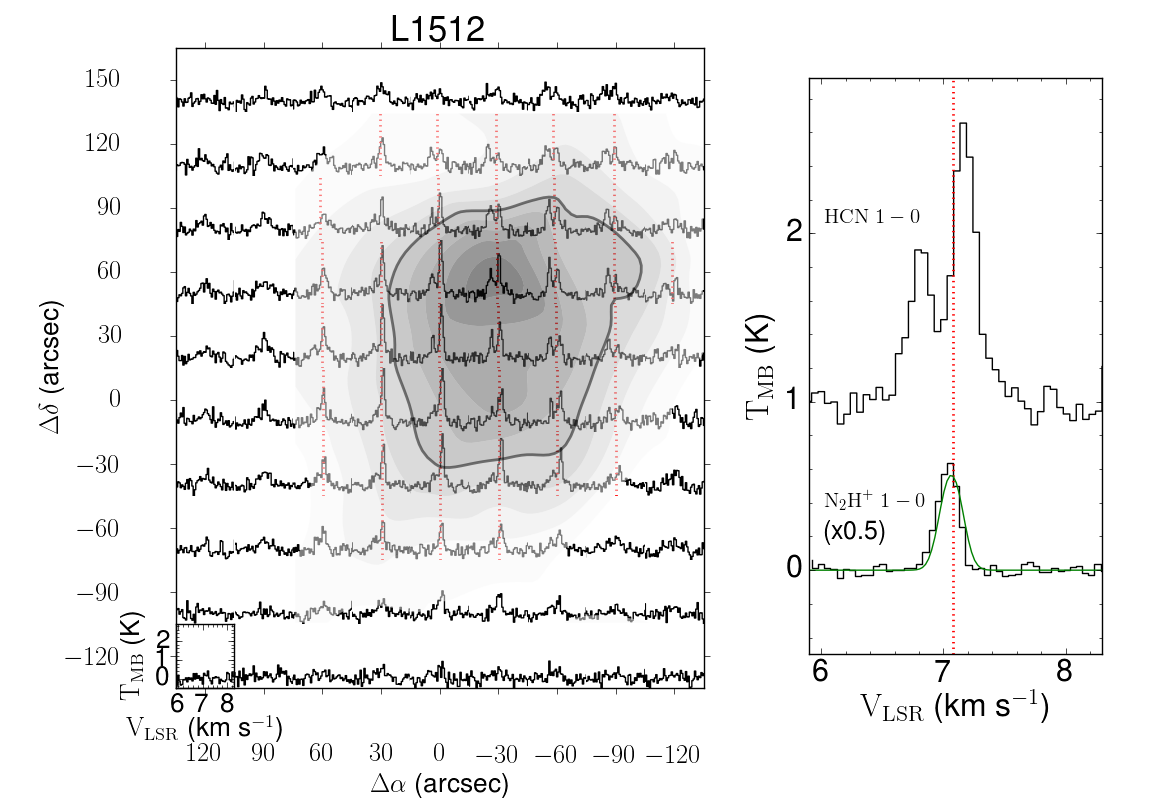} 
\end{array}$
\caption{Same as Figure \ref{fig:HCNmap1} but of three starless cores, CB22, L1517B, and L1512. The right panel displays the average spectra over the half maximum contour of the N$_2$H$^+$ integrated intensity. The minimum level, maximum level, and interval of the N$_2$H$^+$ integrated intensity are as follows; CB22: 0.24, 2.39, 1.19 K km s$^{-1}$, L1517B: 0.32, 3.17, 1.59 K km s$^{-1}$, L1512: 0.41, 4.06, 2.03 K km s$^{-1}$.
\label{fig:HCNmap2}}
\end{figure}

\begin{figure}
\center
$\begin{array}{c}
\includegraphics[bb=0 0 913 800,scale=.22]{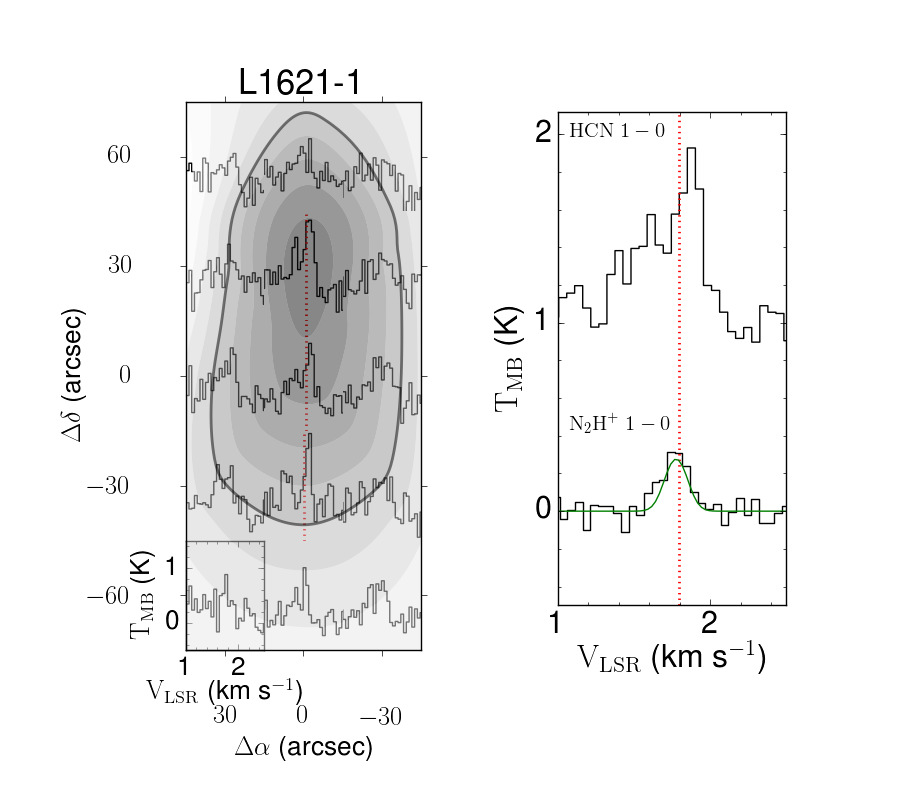} \\
\includegraphics[bb=0 0 1280 800,scale=.22]{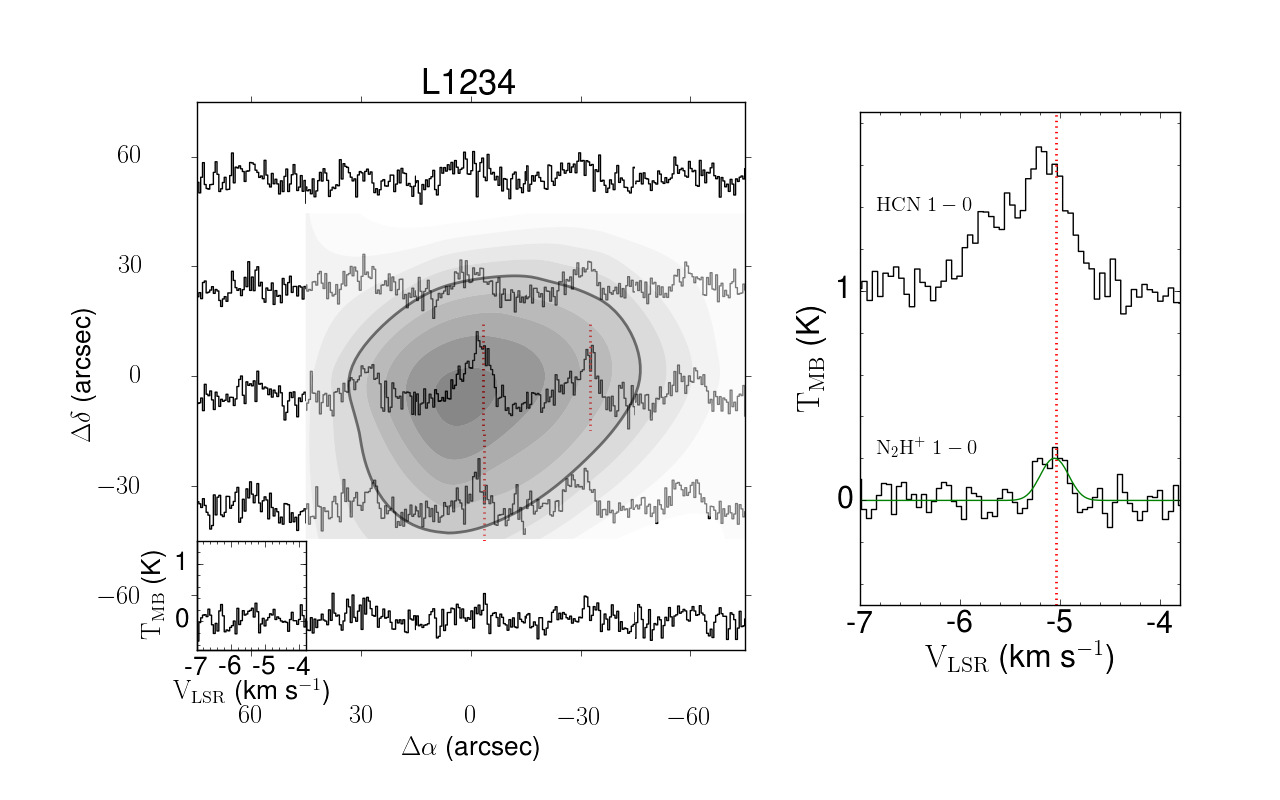} \\
\includegraphics[bb=0 0 1172 800,scale=.22]{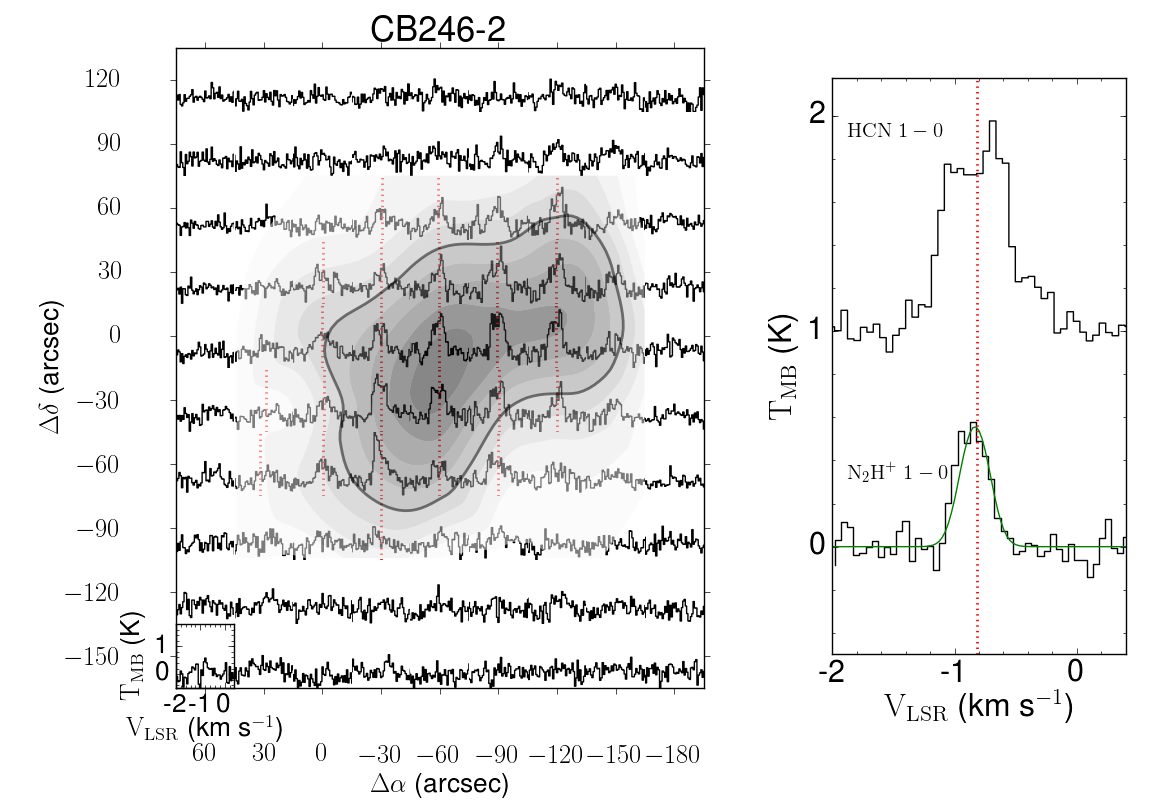} 
\end{array}$
\caption{Same as Figure \ref{fig:HCNmap1} but of three starless cores, L1621-1, L1234, and CB246-2. The right panel displays the average spectra over the half maximum contour of the N$_2$H$^+$ integrated intensity. The minimum level, maximum level, and interval of the N$_2$H$^+$ integrated intensity are as follows; L1621-1: 0.07, 0.68, 0.34 K km s$^{-1}$, L1234: 0.1, 1.0, 0.48 K km s$^{-1}$, CB246-2: 0.26, 2.57, 1.29 K km s$^{-1}$.
\label{fig:HCNmap3}}
\end{figure}

\setcounter{figure}{7} 
\renewcommand{\thefigure}{\arabic{figure}}

\begin{figure}
\includegraphics[bb=0 0 658 321,scale=.7]{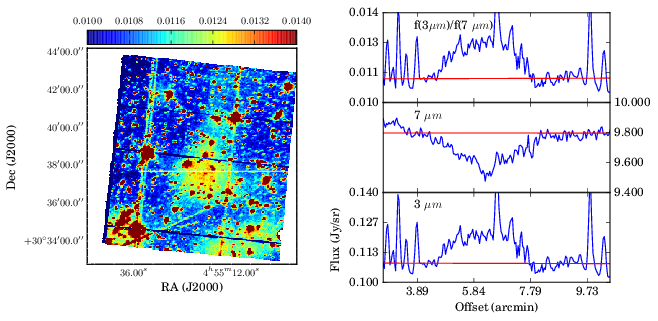} 
\caption{The flux density ratio (f(3 {\micron}) and f(7 {\micron})) map and the cut-profiles (at 3 {\micron} and 7 {\micron} intensity maps) of L1517B as an example of the core shine effect. \label{fig:cutprofile1}}
\end{figure}

\end{document}